\documentclass[12pt]{iopart}

\usepackage{graphicx,amssymb,setstack,latexsym,cite}

\def\<{\langle}
\def\>{\rangle}
\def\tm{$^{\rm (tm)}$ }
\newcommand{\PA}{{\it Physica}}
\newcommand{\NAT}{{\it Nature}}
\newcommand{\EPLOLD}{{\it Europhys.~Lett.~}}
\newcommand{\EPL}{{\it EPL~}}
\newcommand{\JSTAT}{{\it J.~Stat.~Mech.}}

\begin{document}
\title[Critical Casimir forces and adsorption profiles]{Critical Casimir forces and adsorption profiles in the presence of a chemically structured substrate}
\author{Francesco Parisen Toldin and S Dietrich}
\address{Max-Planck-Institut f\"ur Metallforschung,
Heisenbergstr. 3, D-70569 Stuttgart, Germany
}\address{
Institut f\"ur Theoretische und Angewandte Physik,
Universit\"at Stuttgart,
Pfaffenwaldring 57, D-70569 Stuttgart, Germany
}
\eads{\mailto{parisen@mf.mpg.de} and \mailto{dietrich@mf.mpg.de}}

\begin{abstract}
Motivated by recent experiments with confined binary liquid mixtures near demixing, we study the universal critical properties of a system, which belongs to the Ising universality class, in the film geometry. We employ periodic boundary conditions in the two lateral directions and fixed boundary conditions on the two confining surfaces, such that one of them has a spatially homogeneous adsorption preference while the other one exhibits a laterally alternating adsorption preference, resembling locally a single chemical step. By means of Monte Carlo simulations of an improved Hamiltonian, so that the leading scaling corrections are suppressed, numerical integration, and finite-size scaling analysis we determine the critical Casimir force and its universal scaling function for various values of the aspect ratio of the film. In the limit of a vanishing aspect ratio the critical Casimir force of this system reduces to the mean value of the critical Casimir force for laterally homogeneous $++$ and $+-$ boundary conditions, corresponding to the surface spins on the two surfaces being fixed to equal and opposite values, respectively. We show that the universal scaling function of the critical Casimir force for small but finite aspect ratios displays a linear dependence on the aspect ratio which is solely due to the presence of the lateral inhomogeneity. We also analyze the order-parameter profiles at criticality and their universal scaling function which allows us to probe theoretical predictions and to compare with experimental data.
\end{abstract}


\maketitle

\section{Introduction}
\label{sec:intro}
If a fluctuating field is confined between surfaces, effective forces arise between them. In a system close to a second-order phase transition, the order parameter develops such long-ranged fluctuations. The resulting effective force is known as the critical Casimir force. This phenomena, first predicted by Fisher and de~Gennes\cite{FG-78} is the analogue of the Casimir effect in quantum electrodynamics \cite{Casimir-48}.
Reference \cite{Gambassi-09} provides a recent review which illustrates analogies as well as differences between the two effects and guides the reader towards further reviews of the subject and the pertinent original literature.

The critical Casimir force is characterized by a universal scaling function, which is determined by the bulk and surface universality classes (UC) \cite{Binder-83,Diehl-86} of the confined system. It is independent of microscopic details of the system, and it depends only on a few global and general properties, such as the spatial dimension $d$, the number of components of the order parameter, the shape of the confinement, and the boundary conditions (b.c.) there \cite{Krech-94,Krech-99,BTD-00}.
We are solely interested in isotropic systems because the critical Casimir force is only active in fluid systems which allow the ordering degrees of freedom to move in and out of the system; this is not the case for magnets for which the critical Casimir force is not a measureable quantity. (For a discussion of the potential relevance of lattice anisotropies with respect to universality \cite{Dohm-08,KD-09} see Ref.~\cite{DC-09}.)

In recent years the critical Casimir effect has attracted numerous experimental \cite{he4,cwetting,qwetting,HHGDB-08,GMHNHBD-09,SZHHB-08,NHB-09} and even more theoretical investigations. Critical Casimir forces can be obtained indirectly by studying wetting layers of fluids close to a critical end point \cite{NI-85,KD-92b}. In this context, $^4$He wetting films close to the onset of superfluidity \cite{he4} and wetting layers of classical \cite{cwetting} and quantum \cite{qwetting} binary liquid mixtures have been studied experimentally.
Only recently direct measurements of the critical Casimir force have been reported \cite{HHGDB-08,GMHNHBD-09,SZHHB-08,NHB-09} by monitoring individual colloidal particles immersed into a binary liquid mixture close to its critical demixing point and exposed to a planar wall.

Until recently theoretical investigations of the Casimir force used, to a large extent, field-theoretical methods. The geometry studied most is the film geometry in the presence of symmetry-conserving boundary conditions, which has been studied within mean field theory, the $\varepsilon-$expansion, and the perturbative expansion at fixed dimension $d=3$ \cite{KD-91,KD-92,Krech-97,CD-98,Uchida-01,CD-01,DK-04,ZRK-04,MZ-05,MD-05,DGS-06,SD-08,Dean-08,Dohm-09,DG-09,KD-09}, real-space renormalization-group theory \protect\cite{NI-85}, and within the large-$N$ expansion \cite{CD-01,DK-04,CDT-97,Dantchev-98,CD-04,DDG-05,DG-08,CHG-09}. Film geometries with symmetry-breaking surface fields have been considered within mean field theory \cite{Krech-97,DSD-07,MMD-10} and by using a local density functional method \cite{BU-08}. In dimension two exact results are available for periodic b.c. \cite{Cardy-87}, symmetry-conserving b.c. \cite{ZRSA-10} and for symmetry-breaking b.c. \cite{AM-09}. The film geometry in the presence of inhomogeneous b.c. (i.e., with patterned substrates) has been investigated within mean field theory \cite{SSD-06} and within Gaussian approximation \cite{KPHSD-04}. Curved boundaries, which are of particular experimental interest, pose additional difficulties. Spherical geometry has been studied within the $\varepsilon-$expansion at criticality \cite{ER-95} and, off criticality, within mean field theory, the small-sphere expansion and the Derjaguin approximation \cite{HSED-98,SHD-03,TKGHD-09}. The critical Casimir force has been studied also for geometrically structured confinements \cite{PD-04,THD-08} and for ellipsoidal colloids \cite{NO-09,KHD-09} within mean field theory.

Early numerical simulations for the critical Casimir force have been employed in Ref.~\cite{Krech-97} for laterally homogeneous b.c.. Quantitatively reliable determinations of critical Casimir forces for laterally homogeneous b.c. have been obtained recently by means of Monte Carlo simulations. In this context, the $XY$ UC \cite{DK-04,Hucht-07,VGMD-07,VGMD-08,Hasenbusch-09b,Hasenbusch-09c,Hasenbusch-09d}, which describes the critical properties of the superfluid phase transition in $^4$He, as well as the Ising UC \cite{DK-04,VGMD-07,VGMD-08,Hasenbusch-10c}, which describes the demixing transition in a binary liquid mixture, have been investigated and the critical Casimir force has been determined.

Experiments with binary liquid mixtures have also been used to study critical Casimir forces acting on a colloid close to a chemically structured substrate \cite{SZHHB-08,NHB-09}, creating a laterally varying adsorption preference. Such a system has been investigated theoretically for the film geometry within mean field theory \cite{SSD-06} and for a sphere near a planar wall within the Derjaguin approximation \cite{TKGHD-09}.
Motivated by these experimental results, and in the view of the lack of bona fide theoretical data in $d=3$ for laterally inhomogeneous critical systems, here we present a Monte Carlo study of a three-dimensional lattice model in the film geometry, representing the Ising UC. We employ periodic boundary conditions in the two lateral directions and fixed boundary conditions on the two confining surfaces. The Ising spins on the upper surface are fixed to $+1$. The lower surface is divided into two halves, one with spins fixed to $-1$ and the other with spins fixed to $+1$, such that the system remains translationally invariant in one lateral direction. The lattice constant of the simple cubic lattice is set to $1$. Locally this mimics a single chemical step as the simplest building element for lateral heterogeneities. However, the presence of lateral periodic b.c. in the direction normal to the chemical step generates an additional chemical step at the lateral boundaries, resulting in a pair of individual chemical steps. In a system of finite lateral size, the presence of these chemical steps gives rise to a line contribution to the free energy. We note that in a system with a finite lateral size these line contributions cannot be due to a single chemical step only. For instance, in the presence of open lateral b.c., besides the single chemical step there is an additional line contribution to the free energy due to the free edges. Our choice of laterally periodic b.c. corresponds to the simplest implementation of a model with a laterally varying adsorption preference. Furthermore, here we consider also the extrapolation to an infinite lateral extent; in this limit, the lateral b.c. are irrelevant and we recover locally a single chemical step. For this system we determine the critical Casimir force and the order-parameter profiles. Certain preliminary results for this kind of system have been reported in Ref.~\cite{FPD-10}.

Taking corrections-to-scaling into account is important in order to be able to extrapolate data for systems of finite size $L$ to the thermodynamic limit $L\rightarrow\infty$. In particular, in the standard three-dimensional Ising model, scaling corrections are proportional to $L^{-\omega}$, with $\omega=0.832(6)$ \cite{Hasenbusch-10}. The presence of non-periodic boundary conditions such as in the film geometry gives rise to additional scaling corrections, the leading ones being proportional to $L^{-1}$, which are numerically difficult to disentangle from the previous ones. In order to avoid the simultaneous presence of these competing corrections, we have studied a so-called improved model \cite{PV-02}, for which the leading scaling corrections $\propto L^{-\omega}$ are suppressed for all observables so that the correction $\propto L^{-1}$ becomes the leading one.

This paper is organized as follows. In \sref{sec:fss} we recall the finite-size scaling theory for a system with non-periodic boundary conditions, paying special attention to the corrections-to-scaling. In doing so we consider the film geometry described above, as well as more general b.c..
In \sref{sec:simulations} we describe the method we use, which is based on a combination of Monte Carlo simulations and numerical integration.
Our results for the Casimir force amplitude at the critical temperature and for the corresponding universal scaling function are presented in \sref{sec:critical} and \sref{sec:theta}, respectively.
In \sref{sec:profiles} we analyze the order-parameter profiles. We summarize our results in \sref{sec:conclusions}.
In order to estimate non-universal amplitudes, the analysis of the high-temperature series of the improved model is presented in \ref{sec:ht}.
In \ref{sec:mc} we give some technical details concerning our Monte Carlo simulations.

\section{Finite-size scaling}
\label{sec:fss}
\subsection{General properties}
\label{sec:fss:general}
We consider a three-dimensional system in a box of size $L \times L_\parallel \times L_\parallel$, which in the thermodynamic limit exhibits a second-order phase transition at the temperature $T=T_c$. In this subsection we provide those finite-size scaling (FSS) properties which hold independently of the choice of the boundary conditions (b.c.), leaving the discussion of their influence to \sref{sec:fss:bc}.
Away from criticality (i.e.,  $L$, $L_\parallel \gg \xi\sim |t|^{-\nu}$, with $\xi$ as the bulk correlation length, $t\equiv (T-T_c)/T_c$, and $\nu$ the thermal critical exponent), the free energy density $\cal F$ per $k_BT$ of the system (i.e., the free energy divided by $LL_\parallel^2k_BT$) can be expanded into specific geometric contributions, corresponding to the bulk, surfaces, lines, and corners \cite{Privman-89}. All these terms but the bulk one depend on the b.c. (see the discussion in the following subsection).

In the critical region and in the presence of an external bulk field $H$, the free energy density can be decomposed into a singular contribution and a non-singular background term:
\begin{equation}
{\cal F}(t,H,L,L_\parallel) = {\cal F}^{\rm (s)}(t,H,L,L_\parallel) + {\cal F}^{\rm (ns)}(t,H,L,L_\parallel).
\label{free_sns}
\end{equation}

As we shall discuss in \sref{sec:fss:bc}, analogously to the expansion away from criticality the non-singular background ${\cal F}^{\rm (ns)}(t,H,L,L_\parallel)$ can be decomposed into specific geometric terms.
According to renormalization-group (RG) theory \cite{Wegner-76}, neglecting corrections to scaling, in spatial dimension $d$ the singular part of the free energy density obeys the following scaling property:
\begin{equation}
\label{free_full_fss}
{\cal F}^{\rm (s)}(t,H,L,L_\parallel) = \frac{1}{L^d}f\left(u_tL^{y_t},u_h L^{y_h},\rho\right),
\end{equation}
where $y_t$ and $y_h$ are the RG-dimensions of the non-linear scaling fields $u_t$ and $u_h$ associated with the deviations from the critical temperature ($u_t\sim t$) and with the external field ($u_h\sim H$), respectively, and $\rho\equiv L/L_\parallel$ is the aspect-ratio. The RG-dimensions $y_t$ and $y_h$ are related to standard critical exponents by $y_t=1/\nu$ and $y_h=(d+2-\eta)/2$. 
The scaling fields are analytical functions of the parameters of the Hamiltonian \cite{Wegner-76,AF-83} which can be expanded as
\begin{equation}
\label{exp_ut}
u_t=a_0t+o(t,H),
\end{equation}
and
\begin{equation}
\label{exp_uh}
u_h=b_0H+o(t,H),
\end{equation}
where $a_0$ and $b_0$ are non-universal constants. Close to the bulk critical point, the bulk correlation length $\xi$ varies as
\begin{equation}
\label{xi_crit}
\xi(t\rightarrow 0^\pm)=\xi_0^\pm|t|^{-\nu},
\end{equation}
where the signs $\pm$ apply for $T\gtrless T_c$ and $\xi_0^\pm$ are non-universal amplitudes forming the universal ratio $\xi_0^+/\xi_0^-=1.896(10)$ \cite{CPRV-02}. Here and in the following $\xi$ is the so-called exponential (or true) correlation length, which governs the exponential decay of the two-point correlation function (see \ref{sec:ht}). Furthermore, one can fix $a_0$ as
\begin{equation}
\label{xivsut}
a_0=\left(\xi_0^+\right)^{-1/\nu},
\end{equation}
so that the scaling variable is given by
\begin{equation}
\tau \equiv u_tL^{y_t}\simeq t \left(\frac{L}{\xi_0^+}\right)^{\frac{1}{\nu}} = ({\rm sign}\ t) \left(\frac{\xi_0^\pm}{\xi_0^+}\right)^{\frac{1}{\nu}}\left(\frac{L}{\xi}\right)^{\frac{1}{\nu}}.
\end{equation}
The scaling function $f$ introduced in eq.~(\ref{free_full_fss}) is expected to be smooth and universal, once the non-universal amplitudes $a_0$ and $b_0$ are fixed.

\subsection{Influence of the boundary conditions}
\label{sec:fss:bc}
We now discuss those properties which do depend on the b.c. of the system. Extending the corresponding specific discussion in \sref{sec:intro}, here we shall consider the following b.c. \footnote{Here we shall not consider geometries with corners. Their presence gives rise to additional logarithmic singularities $\propto \ln L$ in the free energy; see Ref.~\protect\cite{Privman-89} and references therein for a discussion.}:
\begin{itemize}
\item {\tt SP}: semi-periodic b.c., i.e., periodic b.c. in both directions corresponding to the length $L_\parallel$ and various b.c. in the direction corresponding to the length $L$,
\item {\tt FP}: fully periodic b.c..
\end{itemize}

In the case of {\tt SP} b.c. two surfaces $L_\parallel\times L_\parallel$ are present for which we shall consider either free or fixed b.c.\footnote{Fixed boundary conditions can be obtained by applying infinitely strong surface fields. Here we do not address the dependence on scaling variables generated by the presence of {\it finite} surface magnetic fields, which results in interesting crossover phenomena \protect\cite{MMD-10,AM-09}.}. Unless explicitly stated otherwise, the results presented in the following hold for both homogeneous and inhomogeneous b.c..

As we mentioned in \sref{sec:fss:general}, away from criticality the free energy density can be expanded into specific geometric contributions. For {\tt SP} b.c. with laterally {\it homogeneous} (i.e., translationally invariant) b.c. on the surfaces and away from criticality, the free energy density $\cal F$ has the following expansion as a function of the reduced temperature $t$ and the external bulk field $H$ \cite{GMHNHBD-09,KD-92}:
\begin{equation}
{\cal F}(t,H,L\rightarrow\infty,L_\parallel\rightarrow\infty) = f_{\rm bulk}(t,H) + \frac{1}{L} f_{\rm surf}(t,H) + O(e^{-L/\xi}/L),
\label{free_away}
\end{equation}
where $f_{\rm bulk}(t,H)$ is the bulk free energy density in the thermodynamic limit whereas $f_{\rm surf}(t,H)$ is the surface free energy density associated with the surfaces of area $L_\parallel\times L_\parallel$ and $L_\parallel\rightarrow\infty$. If the b.c. on the two surfaces are not translationally invariant in lateral directions, additional terms in the expansion of eq.~(\ref{free_away}) arise. In particular, in the presence of stripes with alternating b.c. an additional contribution $\propto 1/(LL_\parallel)=\rho/L^2$ appears, which corresponds to a line free energy.
In the case of {\tt FP} b.c. $f_{\rm surf}=0$ so that the free energy is simply proportional to the volume $LL_\parallel^2$, with exponentially small corrections. As anticipated in \sref{sec:fss:general}, in eq.~(\ref{free_away}) $f_{\rm bulk}$ is independent of b.c. whereas $f_{\rm surf}$ depends on the b.c. and on the local geometry, here taken to be planar. This allows one to identify a surface free energy for all types of surfaces, and the quantity $f_{\rm surf}(t,H)$ in eq.~(\ref{free_away}) is the sum of the two independent surface free energy densities associated with the confining walls of the corresponding semi-infinite systems \cite{Diehl-86,Privman-89}.

In the critical region the free energy is decomposed into singular and non-singular contributions (see eq.~(\ref{free_sns})).
As in eq.~(\ref{free_away}), the non-singular background ${\cal F}^{\rm (ns)}(t,H,L,\rho)$ can be decomposed further into geometric contributions

\begin{equation}
\fl {\cal F}^{\rm (ns)}(t,H,L\rightarrow\infty,L_\parallel\rightarrow\infty) = f^{\rm (ns)}_{\rm bulk}(t,H) + \frac{1}{L} f^{\rm (ns)}_{{\rm surf}}(t,H) + O(e^{-L}/L),
\label{free_ns}
\end{equation}

where the same considerations concerning the dependence on geometry and boundary conditions apply as explained after eq.~(\ref{free_away}). In eq.~(\ref{free_ns}) the correction terms are characterized by a decay length of the order of the lattice constant, i.e., $1$. In the following we shall neglect such corrections.

Finally, also the universal scaling function $f$, which we have introduced in eq.~(\ref{free_full_fss}), depends on the specific b.c. of the system.

\subsection{Bulk, surface, and excess free energies}
\label{sec:fss:energies}
Off criticality, the expansion in eq.~(\ref{free_away}) applies in that form if the size of the system is large compared with the correlation length, i.e., in the limit $L,L_\parallel\rightarrow\infty$ at fixed $t\neq 0$. This expansion defines the functions $f_{\rm bulk}(t,H)$ and $f_{\rm surf}(t,H)$. Once this limit is taken, one can consider the limit $|t|\rightarrow 0$ \cite{Privman-89}.
On the other hand, off criticality the infinite-volume limit of the free-energy density can be calculated from the functions appearing in eqs.~(\ref{free_sns}-\ref{free_full_fss}): in this limit the scaling variable $tL^{y_t}$ diverges.
In the absence of an external field $H$, the existence of a finite limiting value $f_{\rm bulk}(t,H=0)$ of the bulk free energy density requires that \cite{Privman-88,Eisenriegler-85}
\begin{equation}
\label{asym_fbulk}
f\left(x, 0, \rho\right) \simeq q_\pm|x|^{d\nu},\quad |x|\rightarrow\infty,\ \rho < \infty,
\end{equation}
with universal amplitudes $q_\pm$ and with the signs $\pm$ corresponding to $t\gtrless 0$ so that the bulk free energy density decomposes into non-singular and singular terms:
\begin{equation}
\label{bulk_sns}
{\cal F}(t\rightarrow 0,H=0,L\rightarrow\infty,L_\parallel\rightarrow\infty)= f^{\rm (ns)}_{\rm bulk}(t) + f^{\rm (s)}_{\rm bulk}(t) = f_{\rm bulk}(t),
\end{equation}
with the order of the limits given by $L_\parallel\rightarrow\infty$, $L\rightarrow\infty$, and $t\rightarrow 0$. The singular part of the bulk free energy density is then given by

\begin{eqnarray}
 f^{\rm (s)}_{\rm bulk}(t\rightarrow 0,H=0) &= {\cal F}^{\rm (s)}(t,H=0,L\rightarrow\infty,L_\parallel\rightarrow\infty) \nonumber\\
&= f^{\rm (s)}_{\rm bulk}(t) = q_\pm|t|^{d\nu} = q_\pm|t|^{2-\alpha}.
\label{wegner_bulk}
\end{eqnarray}

As we mention at the end of \sref{sec:fss:bc}, the universal scaling function $f$ depends on the b.c.. However, the bulk free energy density is independent of the b.c.. This implies that the asymptotic behaviour of $f$ reported in eq.~(\ref{asym_fbulk}) is independent of the b.c..
Similarly, in the limit $t\rightarrow 0$ the surface free energy density decomposes into non-singular and singular terms:
\begin{equation}
f_{\rm surf}(t\rightarrow 0,H=0) = f^{\rm (ns)}_{\rm surf}(t\rightarrow 0,H=0) + f^{\rm (s)}_{\rm surf}(t\rightarrow 0,H=0) 
\end{equation}
and the singular part of the surface free energies $f^{\rm (s)}_{\rm surf}(t)$ is obtained by requiring that (compare eqs.~(\ref{free_away}) and (\ref{free_sns})) the following limit exists:
\begin{equation}
\label{limit_surf}
\lim_{L,L_\parallel\rightarrow\infty,\rho<\infty} L\left({\cal F}^{\rm (s)}(t,H=0,L,L_\parallel)-f^{\rm (s)}_{\rm bulk}(t)\right)=f^{\rm (s)}_{\rm surf}(t).
\end{equation}
This in turn requires that
\begin{equation}
\label{partial_asymp}
f\left(x\rightarrow\infty, \rho<\infty\right) - q_\pm|x|^{d\nu} =  r_\pm|x|^{\nu(d-1)} =  r_\pm|x|^{2-\alpha_S},
\end{equation}
with $\alpha_S=\alpha+\nu$ and universal amplitudes $r_\pm$, so that the singular part of the free energy density becomes
\begin{equation}
\label{free_away_wegner}
{\cal F}^{\rm (s)}(t,H=0,L\rightarrow\infty,L_\parallel\rightarrow\infty) = f^{\rm (s)}_{\rm bulk}(t) + \frac{1}{L}f^{\rm (s)}_{\rm surf}(t),
\end{equation}
with
\begin{equation}
\label{wegner_surf}
f^{\rm (s)}_{\rm surf}(t)=r_\pm|t|^{\nu(d-1)},
\end{equation}
which provides the leading correction to eq.~(\ref{wegner_bulk}).
Since for {\tt FP} b.c. there are no surfaces present, in eq.~(\ref{partial_asymp}) and eq.~(\ref{wegner_surf}) one has $r_\pm=0$ and the singular part of the free energy density, in the large-volume limit, behaves as in eq.~(\ref{wegner_bulk}). For {\tt SP} b.c. $r_\pm$ is the sum of the contributions from the two confining surfaces in the corresponding semi-infinite geometries. If one of these two surface exhibits a pair of individual chemical steps, its contribution is the mean value of those amplitudes which belong to those laterally homogeneous surfaces from which the chemical steps are made of.

According to eq.~(\ref{wegner_surf}) the singular surface free energy density exhibits a singularity at the same temperature as the bulk free energy density. Concerning the critical behaviour at the surfaces, one has to distinguish different surface universality classes, which depend on the interactions at the surfaces. In particular, for the critical behaviour of an Ising system, the surface universality class depends on the so-called surface enhancement $c_0$, the sign of which indicates whether the order parameter at the surface is enhanced or diminished compared with its bulk value. The {\it ordinary} universality class holds for systems in which the interactions between the surface spins depress the order parameter and it corresponds to $c_0>0$. The {\it extraordinary} universality class holds if these surface interactions enhance the order parameter and it corresponds to $c_0<0$. In this case the surface orders at a temperature higher than the bulk critical temperature. Still the bulk phase transition in the presence of an already ordered surface region results in a singularity of the surface free energy at $T_c$. The case $c_0=0$ corresponds to the {\it special} universality class which belongs to a multicritical point\footnote{The special transition is located at $c_0=0$ only within mean field theory; the fluctuations shift this point \protect\cite{Diehl-86}.}. In the presence of surface fields, which explicitly break the ${\mathbb Z}_2$ symmetry, even in the absence of an additional bulk external field, the surface is always ordered. This corresponds to the so-called {\it normal} universality class which exhibits the same kind of singularities as the extraordinary universality class \cite{DS-93,BD-04}. Finally, we note that {\it fixed} b.c. can be realized by infinitely strong surface fields, and thus they correspond to the normal/extraordinary surface universality class. We refer to Refs.~\cite{Binder-83,Diehl-86} for more detailed discussions of surface critical phenomena.

The decomposition according to eq.~(\ref{free_away}) and eq.~(\ref{free_away_wegner}) provides a transparent interpretation only for $\xi\ll L$. In the critical region such a decomposition becomes blurred and the scaling function $f$ in eq.~(\ref{free_full_fss}) encodes all geometrical informations regarding the presence of surfaces, edges, etc. Under such circumstances it is possible to \emph{formally} define a scaling function associated with the surfaces by comparing two system, with identical critical bulk behaviour but different b.c. \cite{Mon-89}. To this end we consider a system with either {\tt FP} or {\tt SP} with homogeneous surfaces b.c.; according to eq.~(\ref{free_away}), the free energy densities away from criticality are given by
\begin{equation}
{\cal F}_{\rm \tt FP}(t,H,L,L_\parallel) = f_{\rm bulk}(t,H) + O(e^{-L/\xi}/L),
\end{equation}
and
\begin{equation}
{\cal F}_{\rm \tt SP}(t,H,L,L_\parallel) = f_{\rm bulk}(t,H) + \frac{1}{L} f_{\rm surf}(t,H) + O(e^{-L/\xi}/L),
\end{equation}
where $f_{\rm bulk}(t,H)$ and $f_{\rm surf}(t,H)$ are independent of $L$ and $L_\parallel$ (compare eqs.~(\ref{wegner_bulk}) and (\ref{wegner_surf}), respectively). We now introduce a new quantity
\begin{equation}
\label{surf_fp_sp}
\hat{f}_{\rm surf}(t,H,L,L_\parallel)\equiv L\left({\cal F}_{\rm \tt SP}(t,H,L,L_\parallel)-{\cal F}_{\rm \tt FP}(t,H,L,L_\parallel)\right),
\end{equation}
such that
\begin{equation}
\label{surf_fp_sp_limit}
\hat{f}_{\rm surf}(t,H,L\rightarrow\infty,L_\parallel\rightarrow\infty)=f_{\rm surf}(t,H) + O(e^{-L/\xi}).
\end{equation}
By using eq.~(\ref{free_ns}) and eq.~(\ref{free_full_fss}) the singular part of $\hat{f}_{\rm surf}(t,H,L,L_\parallel)$ can be expressed as
\begin{equation}
\label{surf_formal}
\hat{f}^{\rm (s)}_{\rm surf}(t,H,L,L_\parallel) = \frac{1}{L^2}\Bigg[f_{\rm \tt SP}\left(u_tL^{y_t},u_h L^{y_h},\rho\right) - f_{\rm \tt FP}\left(u_tL^{y_t},u_h L^{y_h},\rho\right)\Bigg],
\end{equation}
where $f_{\rm \tt FP}$ and $f_{\rm \tt SP}$ are the free energy scaling functions of ${\cal F}^{(s)}$ for {\tt FP} and {\tt SP} boundary conditions, respectively (see eq.~(\ref{free_full_fss})). In contrast to the surface free energy density $f^{\rm (s)}_{\rm surf}(t,H)$ in eqs.~(\ref{free_away}) and (\ref{limit_surf}), the extended surface free energy density $\hat{f}^{\rm (s)}_{\rm surf}(t,H,L,L_\parallel)$ introduced in eq.~(\ref{surf_fp_sp}) depends not only on the local geometry but on the whole original reference geometry (only $\lim_{L,L_\parallel\rightarrow\infty}\hat{f}^{\rm (s)}_{\rm surf}$ reduces to $f^{\rm (s)}_{\rm surf}$). In particular for ${\tt SP}$ b.c. $\hat{f}^{\rm (s)}_{\rm surf}$ encodes the presence of {\it both} surfaces, the contributions of which cannot be disentangled, while as mentioned after eq.~(\ref{free_away}), the surface free energy densities $f^{\rm (s)}_{\rm surf}$ and $f^{\rm (ns)}_{\rm surf}$ are the sum of the single-surface contributions in the corresponding semi-infinite geometries. In \sref{sec:theta:step} we shall extend this kind of reasoning in order to include line contributions to the free energy.

Considering for simplicity a system in the absence of an external field $H$, the excess free energy $f_{\rm ex}^{\rm (s)}$ is defined as the remainder of the free energy density ${\cal F}^{\rm (s)}$ after subtraction of the bulk contribution:
\begin{equation}
\label{free_ex_def}
f_{\rm ex}^{\rm (s)}(t,L,L_\parallel)\equiv {\cal F}^{\rm (s)}(t,H=0,L,L_\parallel)-f^{\rm (s)}_{\rm bulk}(t).
\end{equation}
According to eqs.~(\ref{free_full_fss}) and (\ref{wegner_bulk}) it obeys the following scaling law:
\begin{equation}
\label{scaling_excess}
f_{\rm ex}^{\rm (s)}(t,L,L_\parallel) = \frac{1}{L^d}\Delta\left(u_tL^{y_t},\rho\right).
\end{equation}
The critical Casimir force $F_C$ per area $L_\parallel^2$ and per $k_BT$ is defined as
\begin{equation}
\label{casimir_def}
F_C\equiv-\frac{\partial \left(Lf^{\rm (s)}_{\rm ex}\right)}{\partial L}\Big|_{t,L_\parallel}.
\end{equation}
Analogous to eq.~(\ref{free_full_fss}), in $d=3$ the critical Casimir force exhibits the following scaling behaviour:
\begin{equation}
\label{casimir_fss_leading}
F_C\left(t,L,L_\parallel\right)=\frac{1}{L^3}\theta\left(u_tL^{y_t},\rho\right)
\end{equation}
with
\begin{equation}
\theta(\tau,\rho) = (d-1)\Delta(\tau,\rho) - y_t\tau\frac{\partial\Delta(\tau,\rho)}{\partial \tau} - \rho\frac{\partial\Delta(\tau,\rho)}{\partial \rho},
\end{equation}
so that at the critical point $\tau=u_tL^{y_t}=0$
\begin{equation}
\label{theta_Delta}
\theta(0,\rho)=(d-1)\Delta(0,\rho)-\rho\frac{\partial\Delta(0,\rho)}{\partial\rho}\equiv\Theta(\rho).
\end{equation}
We now consider the Taylor series expansion at criticality
\begin{equation}
\Delta(0,\rho)=\sum_{n=0}^\infty \Delta_n \rho^n,
\end{equation}
which does not capture contributions which have a vanishing Taylor expansion, such as $\exp(-L/L_\parallel)=\exp(-1/\rho)$. In this sense eq.~(\ref{theta_Delta}) yields a Taylor expansion for $\Theta(\rho)$:
\begin{equation}
\label{taylor}
\Theta(\rho)=\sum_{n=0}^\infty (d-1-n)\Delta_n \rho^n=\sum_{n=0}^\infty \theta_n \rho^n,
\end{equation}
such that $\theta_{n=d-1}=0$. In $d=3$ this implies that for $\rho\rightarrow 0$ the critical Casimir force {\em at} the critical temperature has no quadratic term $\propto\rho^2$ in its aspect-ratio dependence.

\subsection{Corrections to scaling and improved models}
\label{sec:fss:corrections}
The scaling behaviour discussed in the preceding subsection is valid only up to corrections-to-scaling contributions. We distinguish two types of scaling corrections: non-analytic and analytic ones. The non-analytic corrections are due to the presence of irrelevant operators. In this case in eq.~(\ref{free_full_fss}) an additional dependence on scaling fields arises, which are characterized by negative RG-dimensions. In the finite size scaling limit (FSS), i.e., for $L\rightarrow\infty$, $t\rightarrow 0$ at fixed $\xi/L$, this results in the following expression for the singular part of the free energy density ${\cal F}^{\rm (s)}$:
\begin{equation}
\label{free_full_fss_corrections}
\fl {\cal F}^{\rm (s)}(t,H,L,L_\parallel) = \frac{1}{L^d}\left[f\left(u_tL^{y_t},u_h L^{y_h},\rho\right) + \sum_{i, k\ge 1} L^{ky_i}g_i\left(u_tL^{y_t},u_h L^{y_h},\rho\right)\right],
\end{equation}
where $y_i<0$, $i\ge 1$, are the RG-dimensions of the irrelevant operators and $g_i$ are smooth functions which are universal up to a normalization constant. The leading correction is given by the operator that has the smallest negative dimension. This is usually denoted with $\omega$, so that the leading scaling corrections are $\propto L^{-\omega}$. For the standard three-dimensional Ising model one has $\omega=0.832(6)$ \cite{Hasenbusch-10}.

In a family of models characterized by an irrelevant parameter $\lambda$, it can occur that for a certain choice of $\lambda$ the leading correction-to-scaling term $\propto L^{-\omega}$ vanishes: such a model is called \emph{improved}.
In such models, the observed scaling corrections usually decay much more rapidly: as $L^{-\omega_2}$ with $\omega_2\sim 1.67(11)$ \cite{NR-84} for the three-dimensional Ising universality class. Improved models have turned out to be instrumental for obtaining high-precision results  for critical phenomena \cite{PV-02}. In a lattice model there are also scaling corrections $\propto L^{-\omega_{\rm NR}}$ due to the presence of non rotationally-invariant irrelevant operators, i.e., operators which break the rotational invariance. The leading one for lattices with cubic symmetry leads to $\omega_{\rm NR}\simeq 2$ \cite{CPRV-98}.

Another type of scaling corrections is provided by the so-called analytic scaling corrections, which can stem from various sources. The linear expansion of the scaling fields introduced in eqs.~(\ref{exp_ut}) and (\ref{exp_uh}) is valid only up to higher-order terms in the expansion. For instance, it is easy to see that an additional term $\propto t^2$ in eq.~(\ref{exp_ut}) gives rise, in the FSS limit, to scaling corrections $\propto L^{-1/\nu}$. It is worthwhile to note that, for the Ising universality class and general $O(N)$ models, one has $\nu<1$ so that these corrections are usually not observed because they are subdominant.
Analytic corrections can be also due to boundary conditions: not fully periodic b.c. induce additional corrections, which are proportional to $L^{-1}$. It was first proposed in Ref.~\cite{CF-76} in the context of studying surface susceptibilities, that such scaling corrections can be absorbed by the substitution $L\rightarrow L+c$, where $c$ is a non-universal, temperature--independent length. This property has been recently checked numerically in Refs.~\cite{Hasenbusch-08,Hasenbusch-09,Hasenbusch-09b} for the $XY$ model for {\tt SP} b.c. with free surfaces and in Ref.~\cite{Hasenbusch-10c} for the Ising model for {\tt SP} b.c. with homogeneously fixed surface spins. We note that, in the same spirit of the expansion of eqs.~(\ref{exp_ut}) and (\ref{exp_uh}), one should expect this substitution to be correct up to higher orders in $1/L$, i.e., $L\rightarrow L+c+O(1/L)$. However, such higher-order terms would induce corrections to scaling $\propto L^{-2}$ which interfer with those due to the aforementioned breaking of rotational invariance; accordingly, scaling corrections $\propto L^{-2}$ are captured by the more general ansatz given by eq.~(\ref{free_full_fss_corrections}).

In the present contribution we study the critical Casimir force using an improved model in three spatial dimensions, employing {\tt SP} b.c. with fixed surface spins. On the basis of the above discussion for such a model the leading scaling corrections are expected to be proportional to $L^{-1}$. Furthermore, we assume that also in this case in leading order we can absorb such a scaling correction by the substitution $L\rightarrow L+c$. In this sense eq.~(\ref{casimir_fss_leading}) is replaced by
\begin{equation}
\label{casimir_fss}
F_C\left(t,L,L_\parallel\right)=\frac{1}{(L+c)^3}\theta\left(a_0t(L+c)^{y_t},\frac{L+c}{L_\parallel}\right),
\end{equation}
where we have used the expansion of eq.~(\ref{exp_ut}) to leading order, thus neglecting corrections $\propto L^{-1/\nu}\sim L^{-1.6}$. We have also neglected corrections to scaling due to next-to-leading irrelevant operators, which are expected to scale as $L^{-\omega_2}$, $\omega_2\sim 1.67(11)$ \cite{NR-84}.
If we expand eq.~(\ref{casimir_fss}) for $L\rightarrow\infty$, at fixed $tL^{y_t}$ and $\rho$, we obtain the following expression for the leading behaviour and for the scaling corrections of the critical Casimir force:
\begin{eqnarray}
\label{corr_scaling_casimir}
F_C\left(t,L,L_\parallel\right)=\frac{1}{L^3}\left[\theta\left(a_0tL^{y_t},\rho\right)+\frac{c}{L}\psi\left(a_0tL^{y_t},\rho\right)\right],\nonumber\\
\psi\left(\tau,\rho\right)\equiv-3\theta(\tau,\rho)+\frac{1}{\nu}\tau\frac{\partial\theta\left(\tau,\rho\right)}{\partial \tau}+\rho\frac{\partial\theta\left(\tau,\rho\right)}{\partial \rho}.
\end{eqnarray}
The comparison of eq.~(\ref{corr_scaling_casimir}) with eq.~(\ref{free_full_fss_corrections}) reveals an important difference in the correction-to-scaling terms. While in eq.~(\ref{free_full_fss_corrections}) the functions $g_i$, which provide the scaling corrections, are independent from the leading scaling function $f$, in eq.~(\ref{corr_scaling_casimir}) the correction-to-scaling function $\psi$ follows from the leading scaling function $\theta$. In principle this prediction provides an opportunity to check the appealing ansatz in eq.~(\ref{casimir_fss}).

\subsection{Finite-size scaling of the order parameter}
\label{sec:fss:profiles}
The scaling properties of the order parameter can be determined from a generalization of eq.~(\ref{free_full_fss}) by introducing a spatially varying external field \cite{Binder-83,Diehl-86}. Here we consider {\tt SP} b.c. with fixed surface spins. The system is described by the coordinates $x,y,z$ on the lattice, with $z$ the coordinate corresponding to the length $L$. Furthermore we restrict ourselves to the case in which the system is translationally invariant in the $y$ direction, i.e., the b.c. of fixed surface spins are translationally invariant in one direction, corresponding to the length $L_\parallel$.
The order parameter $\Phi$ (i.e., the magnetization per volume) exhibits, in leading order, the following scaling form:
\begin{equation}
\label{scaling_mxz}
\Phi(t,x,z,L,L_\parallel) = B |t|^\beta P_\pm\left(\frac{x}{\xi_\pm},\frac{z}{\xi_\pm},\frac{L}{\xi_\pm},\rho\right),\qquad \xi_\pm=\xi_0^\pm|t|^{-\nu},
\end{equation}
where $\xi_\pm$ is the correlation length for $t \gtrless 0$ with non-universal amplitudes $\xi_0^\pm$ (see eq.~(\ref{xi_crit})). The scaling functions $P_\pm(\tilde{x},\tilde{z},\tilde{L},\rho)$ (for $t \gtrless 0$) are universal and depend also on the universality classes of the two confining surfaces. $B$ is the non-universal amplitude of the spontaneous bulk magnetization at $T<T_c$:
\begin{equation}
\label{spontm}
\Phi_{\rm bulk}(t\rightarrow 0^-)=B |t|^\beta.
\end{equation}
The amplitude of the scaling functions $P_\pm$ is fixed by the requirement that eq.~(\ref{spontm}) is reproduced in the bulk limit, i.e., $\lim_{\tilde{L}\rightarrow\infty} P_-(\tilde{x},\tilde{z}=\tilde{L}/2,\tilde{L},\rho)=1$. Due to the fixed b.c., in the limits $t\rightarrow 0^\pm$ the scaling functions $P_\pm$ give a unique, nonvanishing order parameter profile, respecting the universality of the ratio $\xi_0^+/\xi_0^-=1.896(10)$ \cite{CPRV-02}.
Equation (\ref{scaling_mxz}) can be rewritten as
\begin{equation}
\label{scaling_mxz2}
\Phi(t,x,z,L,L_\parallel) = B \left(\frac{L}{\xi_0^\pm}\right)^{-\beta/\nu}\phi_\pm\left(\frac{x}{L}, \frac{z}{L},\frac{L}{\xi_\pm},\rho\right),
\end{equation}
where
\begin{equation}
\label{scaling_mxz3}
\phi_\pm(\hat{x},\hat{z},\tilde{L},\rho) = \tilde{L}^{\beta/\nu}P_\pm\left(\tilde{x}=\hat{x}\tilde{L},\tilde{z}=\hat{z}\tilde{L},\tilde{L},\rho\right)
\end{equation}
are also universal scaling functions.

If the order parameter on the whole of each surface is either fixed to $+1$ ($(+)$ b.c. in the following) or $-1$ ($(-)$ b.c.) the system is translationally invariant in both lateral directions. In this case the order parameter does not depend on $x$ and eqs.~(\ref{scaling_mxz}) and (\ref{scaling_mxz2}) reduce to
\begin{eqnarray}
\label{scaling_mz}
&\Phi(t,z,L,L_\parallel) = B |t|^\beta P^{\rm (hom)}_\pm\left(\frac{z}{\xi_\pm},\frac{L}{\xi_\pm},\rho\right),\\
\label{scaling_mz2}
&\Phi(t,z,L,L_\parallel) = B \left(\frac{L}{\xi_0^\pm}\right)^{-\beta/\nu}\phi^{\rm (hom)}_\pm\left(\frac{z}{L},\frac{L}{\xi_\pm},\rho\right),
\end{eqnarray}
with, again, universal scaling functions $P^{\rm (hom)}_\pm(\tilde{z},\tilde{L},\rho)$ and $\phi^{\rm (hom)}_\pm(\hat{z},\tilde{L},\rho) = \tilde{L}^{\beta/\nu}P_\pm^{\rm (hom)}(\tilde{z}=\hat{z}\tilde{L},\tilde{L},\rho)$; the superscript (hom) indicates laterally homogeneous b.c..

For fixed $t$, fixed $z$, fixed aspect ratio $\rho$ ($\rho < \infty$), and for increasing separation $L$ the scaling functions introduced so far converge to the corresponding half-space (semi-infinite) scaling functions for the confining surface located at $z=0$:
\begin{eqnarray}
\label{limit_semiinfty}
&P_\pm(\tilde{x},\tilde{z},\tilde{L}\rightarrow\infty,\rho) \rightarrow P_{\pm,\infty}(\tilde{x},\tilde{z}),\\
\label{limit_semiinfty_hom}
&P^{\rm (hom)}_\pm(\tilde{z},\tilde{L}\rightarrow\infty,\rho) \rightarrow P^{\rm (hom)}_{\pm,\infty}(\tilde{z}),
\end{eqnarray}
where $P_{\pm,\infty}(\tilde{x},\tilde{z})$ and $P^{\rm (hom)}_{\pm,\infty}(\tilde{z})$ describe the order parameter profiles for a semi-infinite ($\infty/2$) geometry in the presence of an inhomogeneous and a homogeneous surface, respectively, at $z=0$:
\begin{eqnarray}
\label{scaling_mz_half}
&\Phi(t,z) = B |t|^\beta P^{\rm (hom)}_{\pm,\infty}\left(\frac{z}{\xi_\pm}\right)\qquad &{\rm hom.~surf.},\quad \infty/2,\\
\label{scaling_mz_half2}
&\Phi(t,x,z) = B |t|^\beta P_{\pm,\infty}\left(\frac{x}{\xi^\pm},\frac{z}{\xi^\pm}\right)\qquad &{\rm inhom.~surf.},\quad \infty/2.
\end{eqnarray}
In the limit $L\rightarrow\infty$ at a fixed aspect ratio $\rho$, one also has $L_\parallel\rightarrow\infty$. Accordingly, the limits in eqs.~(\ref{limit_semiinfty}) and (\ref{limit_semiinfty_hom}) describe a system with an infinite lateral extension. In such a geometry no quantity depends on the original aspect ratio $\rho$ of the confined system. Therefore we have dropped the $\rho$-dependence on the r.h.s. of eqs.~(\ref{limit_semiinfty}) and (\ref{limit_semiinfty_hom}), as well as in eqs.~(\ref{scaling_mz_half}) and (\ref{scaling_mz_half2}).
For homogeneous surfaces in the film geometry, the short-distance behaviour of the scaling functions in eq.~(\ref{scaling_mz}) and (\ref{scaling_mz_half}) is given by \cite{Diehl-86,FD-95}
\begin{eqnarray}
\label{reduction}
&P^{\rm (hom)}_\pm(\tilde{z}\rightarrow 0,\tilde{L},\rho) \simeq c_\pm \tilde{z}^{-\beta/\nu},\\
\label{reduction_infty}
&P^{\rm (hom)}_{\pm,\infty}(\tilde{z}\rightarrow 0) \simeq c_\pm \tilde{z}^{-\beta/\nu},
\end{eqnarray}
i.e., the {\it leading} behaviour near the surface is not influenced by the finite film thickness which, however, gives rise to distant-wall corrections (see below)\footnote{In principle, the r.h.s. of eq.~(\protect\ref{reduction}) could depend on the aspect ratio $\rho$. However, since a confined system displays a finite correlation length and the order parameter is a {\it local} quantity, we expect that the effect of a nonzero aspect ratio $\rho$, i.e., a finite lateral size, will be effectively very weak, if not absent at all. Thus for simplicity in eq.~(\protect\ref{reduction}) we have omitted a possible dependence of the constants $c_\pm$ on $\rho$. The data which will be presented in \protect\sref{sec:profiles} support this observation.}. Equations (\ref{reduction}) and (\ref{reduction_infty}) imply that in both cases at criticality the order parameter varies algebraicly:
\begin{equation}
\label{decay}
\Phi(t=0,z\rightarrow 0,L,L_\parallel) \simeq B c_\pm (z/\xi_0^\pm)^{-\beta/\nu}.
\end{equation}
The amplitudes $c_+$ and $c_-$ are universal \cite{FD-95} with $c_+/c_-=(\xi_0^+/\xi_0^-)^{-\beta/\nu}$. The presence of a second wall induces corrections \cite{EKD-93,ES-94} to the leading decay (eq.~(\ref{decay})) of the critical order parameter. For $L_\parallel\rightarrow\infty$ the leading distant-wall corrections are given by
\begin{equation}
\label{wallcorr}
\Phi_{ab}(z,t=0)=\Phi_{a,\infty}(z,t=0){\cal F}_{ab}(z/L)
\end{equation}
with
\begin{equation}
\label{wallcorr2}
{\cal F}_{ab}(z/L)\equiv 1 - C_a (d-1)\Delta_{ab}\left(\frac{z}{L}\right)^d,
\end{equation}
where $\Phi_{a,b}$ is the order-parameter profile in the presence of walls with surface universality classes $a$ and $b$ and $\Phi_{a,\infty}$ is the semi-infinite order-parameter profile for the near wall, given by eq.~(\ref{scaling_mz_half}). In the universal scaling function ${\cal F}_{ab}(z/L)$ the coefficient $C_a$ depends only on the boundary conditions of the wall $a$, whereas the influence of the boundary condition at the distant wall enters via $\Delta_{ab}=\Delta_{ab}(0,0)$ (eq.~(\ref{scaling_excess})), which is the Casimir amplitude of the free energy at criticality.

The expressions in eqs.~(\ref{scaling_mxz}), (\ref{scaling_mxz2}), (\ref{scaling_mz}), (\ref{scaling_mz2}), (\ref{scaling_mz_half}), and (\ref{scaling_mz_half2}) are valid up to corrections to scaling. In analogy to eqs.~(\ref{free_full_fss_corrections}) and (\ref{casimir_fss}) we expect, to leading order, the occurrence of corrections to scaling $\propto L^{-\omega}$ due to irrelevant operators and of corrections to scaling $\propto L^{-1}$ due to the boundary conditions. Since we shall study an improved model, the latter ones are expected to be the leading corrections to scaling.

\section{Model and method}
\label{sec:simulations}
As discussed in \sref{sec:fss:corrections}, in order to have a better control on the finite-size scaling corrections, we resort to an improved model belonging to the 3D Ising universality class. To this end we chose the Blume-Capel model \cite{Blume-66,Capel-66}. It is defined on a three-dimensional simple cubic lattice, with a spin variable $S_i$ on each site $i$ which can take the values $S_i=-1$, $0$, $1$. The reduced Hamiltonian for nearest neighbour interactions is
\begin{equation}
\label{BC}
{\cal H}=-\beta\sum_{\<i j\>}S_i S_j + D\sum_i S_i^2,\qquad S_i=-1,0,1,
\end{equation}
so that the Gibbs weight is $\exp(-\cal H)$. In line with the convention used in Refs.~\cite{Hasenbusch-10c,CPRV-02,Hasenbusch-01}, in the following we shall keep $D$ constant, considering it as a part of the integration measure over $\{S_i\}$, while we vary the coupling parameter $\beta$, which is proportional to the inverse temperature, $\beta\sim 1/T$. The relation between the dimensionless parameters $\beta$ and $D$ characterizing the reduced Hamiltonian in eq.~(\ref{BC}) and the physical parameters can be obtained by introducing coupling constants $\hat{J}$ and $\hat{D}$, so that the reduced Hamiltonian at temperature $T$ reads
\begin{equation}
\label{BCreal}
{\cal H}=\frac{1}{k_B T}\left(-\hat{J}\sum_{\<i j\>}S_i S_j + \hat{D}\sum_i S_i^2\right),
\end{equation}
where $\hat{J}$, $\hat{D}$, and $k_B T$ have the dimension of energy. The comparison of eqs.~(\ref{BC}) and (\ref{BCreal}) gives $\beta=\hat{J}/(k_BT)$ and $D=\hat{D}/(k_BT)$. In the limit $D\rightarrow -\infty$, one recovers the usual Ising model, because in this limit any state for which there is an $i_0$ such that $S_{i_0}=0$ is suppressed relative to the states $\{S_i=\pm 1\}$. For $d\ge 2$, the model exhibits a phase transition at $\beta_c=\beta_c(D)$ which is second order for $D\le D_{\rm tri}$, and first order for $D>D_{\rm tri}$. The value of $D_{\rm tri}$ in $d=3$ has been determined as $D_{\rm tri}\simeq 2.006$ \cite{Deserno-97} and as $D_{\rm tri}\simeq 2.05$ \cite{HB-98}. At $D=0.641(8)$ \cite{Hasenbusch-01} the model is improved, i.e., leading corrections to scaling $\propto L^{-\omega}$ with $\omega=0.832(6)$ \cite{Hasenbusch-10}\footnote{This paper provides an updated value of the coupling $D=0.656(20)$ for which the model (eq.~(\protect\ref{BC})) is ``improved'' \cite{PV-02}, as well as an updated value of $\beta=0.38567122(5)$ for which the model is critical at $D=0.641$.} are suppressed. At this value of the reduced coupling $D$ the model is critical for $\beta=\beta_c=0.3856717(10)$ \cite{CPRV-02}. We mention that for this improved model, i.e., at $D=0.641$, $25^{\rm th}$-order high-temperature expansion series are available \cite{CPRV-02}. In \ref{sec:ht}, from these series we infer (in units of the lattice constant)
\begin{equation}
\label{xi0}
\xi_0^+=0.415(2)
\end{equation}
as the value of the non-universal amplitude of the true correlation length above $T_c$ (see eq.~(\ref{xi_crit})).

In the following we use as the values of the critical exponents $\nu=0.63012(16)$ and $\eta=0.03639(15)$ which have been obtained by analyzing the $25^{\rm th}$-order high-temperature expansion series for three improved models within the Ising universality class \cite{CPRV-02}.

In order to determine the critical Casimir force, we follow the approach introduced in Ref.~\cite{VGMD-07}, which we briefly describe here. For two reduced Hamiltonian ${\cal H}_1$ and ${\cal H}_2$ associated with the same configuration space $\{C\}$ we construct the convex combination ${\cal H}(\lambda)$
\begin{equation}
\label{crossover_H}
{\cal H}(\lambda) \equiv \left(1-\lambda\right){\cal H}_1 + \lambda{\cal H}_2,\qquad \lambda\in \left[0,1\right].
\end{equation}
This Hamiltonian ${\cal H}(\lambda)$ leads to the free energy ${\mathrm F}(\lambda)$ in units of $k_BT$. Its derivative is
\begin{equation}
\label{free_derivative}
\frac{\partial {\mathrm F}(\lambda)}{\partial\lambda}=\frac{\sum_{\{C\}}\frac{\partial {\cal H}(\lambda)}{\partial\lambda}e^{-{\cal H}(\lambda)}}{\sum_{\{C\}}e^{-{\cal H}(\lambda)}}.
\end{equation}
Combining eqs.~(\ref{crossover_H}) and (\ref{free_derivative}) we can determine the free energy difference as
\begin{equation}
\label{free_diff}
{\mathrm F}(1)-{\mathrm F}(0)=\int_0^1 d\lambda \frac{\partial {\mathrm F}(\lambda)}{\partial\lambda}=\int_0^1 d\lambda \<{\cal H}_2-{\cal H}_1\>_\lambda,
\end{equation}
where $\<{\cal H}_2-{\cal H}_1\>_\lambda$ is the thermal average of the observable ${\cal H}_2-{\cal H}_1$ with the statistical weight $\exp(-{\cal H}(\lambda))$. For every $\lambda$ this average is accessible to standard Monte Carlo simulations. Finally the integral appearing in eq.~(\ref{free_diff}) is performed numerically, yielding the free energy difference between the systems governed by the Hamiltonian ${\cal H}_2$ and ${\cal H}_1$, respectively.

\begin{figure}
\begin{center}
\includegraphics[width=30em,keepaspectratio]{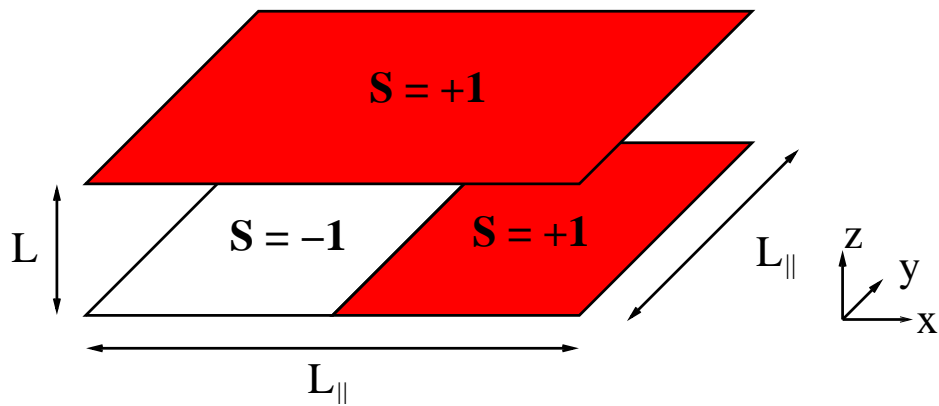}
\end{center}
\caption{Film geometry with aspect ratio $\rho=L/L_\parallel$ confined by a homogeneous upper surface and by a lower surface with a chemical step.}
\label{lattice}
\end{figure}

To be specific, we consider a three-dimensional lattice $L\times L_\parallel\times L_\parallel$ with ${\tt SP}$ b.c. (see \sref{sec:fss:general}), with periodic b.c. in the lateral directions $x$ and $y$, and fixed spins at the two surfaces $z=0$ and $z=L-1$, so that there are $L-2$ layers of fluctuating spins. The spins at the upper surface $z=L-1$ are fixed to $+1$, and on the lower surface $z=0$ we employ a single, straight chemical step, where the surface is divided into two halves, one ($x<0$) with spins fixed to $-1$ and the other half ($x\ge 0$) with spins fixed to $+1$. The presence of lateral periodic b.c. in the $x$ direction generates an additional chemical step at the lateral boundaries, resulting in a system with a pair of individual chemical steps. This geometry is illustrated in \fref{lattice}.

Following Refs.~\cite{VGMD-07,VGMD-08}, we apply eq.~(\ref{free_diff}) with ${\cal H}_1$ as the Hamiltonian of the lattice shown in \fref{lattice} and ${\cal H}_2$ as the Hamiltonian of a lattice $(L-1)\times L_\parallel\times L_\parallel$ plus a separate two-dimensional layer, so that both Hamiltonian share the same configuration space. This layer can be inserted into the film by varying the coupling $(1-\lambda)\beta$ with its neighbouring planes between $0$ and $\beta$. With this we evaluate the following quantity:
\begin{equation}
\label{I_def}
I\left(\beta, L, L_\parallel\right) \equiv \frac{1}{L_\parallel^2}\int_0^1 d\lambda \<{\cal H}_2-{\cal H}_1\>_\lambda.
\end{equation}
In accordance with eqs.~(\ref{free_ns}) and (\ref{free_ex_def}), the free energies (in units of $k_BT$) corresponding to ${\cal H}_1$ and ${\cal H}_2$ can be expressed as
\begin{eqnarray}
\label{F1}
\fl {\mathrm F}_1 = {\mathrm F}(0) = LL_\parallel^2f^{\rm (ns)}_{\rm bulk}(t)+ L_\parallel^2f^{\rm (ns)}_{\rm surf}(t) + L_\parallel f^{\rm (ns)}_{\rm line}(t) +
LL_\parallel^2f^{\rm (s)}_{\rm bulk}(t)+ LL_\parallel^2f^{\rm (s)}_{\rm ex}(t, L, L_\parallel),\\
\fl {\mathrm F}_2 = {\mathrm F}(1) = (L-1)L_\parallel^2f^{\rm (ns)}_{\rm bulk}(t)+ L_\parallel^2f^{\rm (ns)}_{\rm surf}(t) + L_\parallel f^{\rm (ns)}_{\rm line}(t) \nonumber\\
\label{F2}
\qquad + (L-1)L_\parallel^2f^{\rm (s)}_{\rm bulk}(t)+ (L-1)L_\parallel^2f^{\rm (s)}_{\rm ex}(t, L-1, L_\parallel)+ L_\parallel^2f_{2D}(t),
\end{eqnarray}
where $f^{\rm (ns)}_{\rm line}(t)$ represents a (possible) line contribution to the non-singular part of the free energy due to the presence of the pair of individual chemical steps (see the discussion after eq.~(\ref{free_away})) and $f_{2D}(t)$ is the free energy density per area $L_\parallel^2$ and in units of $k_BT$ of the additional 2D layer, which is not critical near the phase transition in  $d=3$ we are interested in. By combining eq.~(\ref{I_def}) with eqs.~(\ref{free_diff}), (\ref{F1}), (\ref{F2}), and (\ref{bulk_sns}) one has
\begin{equation}
\label{I_inter}
I\left(\beta, L, L_\parallel\right) = \left[f_{2D}(t) - f_{\rm bulk}(t)\right] - \left[\bar{f}^{\rm (s)}_{\rm ex}(t, L, L_\parallel)-\bar{f}^{\rm (s)}_{\rm ex}(t, L-1, L_\parallel)\right],
\end{equation}
with
\begin{eqnarray}
f_{\rm bulk}(t) = f^{\rm (ns)}_{\rm bulk}(t) + f^{\rm (s)}_{\rm bulk}(t),\\
\label{fexbar}
\bar{f}^{\rm (s)}_{\rm ex}(t,L,L_\parallel)\equiv Lf^{\rm (s)}_{\rm ex}(t, L, L_\parallel),\\
t=\frac{\beta_c-\beta}{\beta}.
\end{eqnarray}
A Taylor expansion of the last term in eq.~(\ref{I_inter}) around the film thickness $L-1/2$ gives
\begin{equation}
\label{fexbar_taylor}
\fl \bar{f}^{\rm (s)}_{\rm ex}(t,L,L_\parallel)-\bar{f}^{\rm (s)}_{\rm ex}(t,L-1,L_\parallel) = \frac{\partial \bar{f}^{\rm (s)}_{\rm ex}}{\partial L}\left(t,L-\frac{1}{2},L_\parallel\right) + \frac{1}{24} \frac{\partial^3 \bar{f}^{\rm (s)}_{\rm ex}}{\partial L^3}\left(t,L-\frac{1}{2},L_\parallel\right).
\end{equation}
By combining eqs.~(\ref{fexbar}) and (\ref{casimir_def}), we have
\begin{equation}
\label{fexbar_first}
\frac{\partial \bar{f}^{\rm (s)}_{\rm ex}}{\partial L}\left(t,L,L_\parallel\right) = -F_C\left(t,L,L_\parallel\right),
\end{equation}
i.e., the first term in the expansion (\ref{fexbar_taylor}) corresponds to the critical Casimir force. Neglecting for the time being corrections to scaling, we substitute eq.~(\ref{casimir_fss_leading}) in eq.~(\ref{fexbar_first}) obtaining
\begin{equation}
\label{fexbar_first2}
\fl
\frac{\partial \bar{f}^{\rm (s)}_{\rm ex}}{\partial L}\left(t,L,L_\parallel\right) = -\frac{1}{L^3}\theta\left(u_tL^{y_t},\rho\right) = -\frac{1}{L^3}\theta\left(\tau,\rho\right), \qquad \tau=u_tL^{y_t}, \quad y_t=\frac{1}{\nu}.
\end{equation}
 The higher order derivatives follow from eq.~(\ref{fexbar_first2}):
\begin{eqnarray}
\label{fexbar_second}
\fl \frac{\partial^2 \bar{f}^{\rm (s)}_{\rm ex}}{\partial L^2}\left(t,L,L_\parallel\right) = \frac{3}{L^4}\theta\left(\tau,\rho\right) - \frac{y_t}{L^4}\tau\frac{\partial\theta}{\partial\tau}\left(\tau,\rho\right) - \frac{1}{L^4}\rho\frac{\partial\theta}{\partial\rho}\left(\tau,\rho\right) \equiv \frac{1}{L^4}\theta_2\left(\tau,\rho\right),\\
\label{fexbar_third}
\fl \frac{\partial^3 \bar{f}^{\rm (s)}_{\rm ex}}{\partial L^3}\left(t,L,L_\parallel\right) = -\frac{4}{L^5}\theta_2\left(\tau,\rho\right) + \frac{y_t}{L^5}\tau\frac{\partial\theta_2}{\partial\tau}\left(\tau,\rho\right) + \frac{1}{L^5}\rho\frac{\partial\theta_2}{\partial\rho}\left(\tau,\rho\right) \equiv \frac{1}{L^5}\theta_3\left(\tau,\rho\right).
\end{eqnarray}
Inserting eqs.~(\ref{fexbar_first}), (\ref{fexbar_first2}) and (\ref{fexbar_third}) in the expansion (\ref{fexbar_taylor}) we obtain:
\begin{eqnarray}
\fl \bar{f}^{\rm (s)}_{\rm ex}(t,L,L_\parallel)-\bar{f}^{\rm (s)}_{\rm ex}(t,L-1,L_\parallel) \nonumber\\
=-F_C\left(t,L-\frac{1}{2},L_\parallel\right)\left[1 - \frac{1}{24\left(L-1/2\right)^2}\frac{\theta_3\left(u_t\left(L-1/2\right)^{y_t},\rho\right)}{\theta\left(u_t\left(L-1/2\right)^{y_t},\rho\right)}\right].
\label{fexbar_taylor2}
\end{eqnarray}
In the FSS limit, i.e., in the limit $L\rightarrow\infty$ at a fixed ratio $\xi/L$ or fixed $\tau$, the term in brackets represents a scaling correction $\sim L^{-2}$. Such a correction is negligible relative to the leading scaling correction $\sim L^{-1}$ which we expect for the model under study here (see the discussion in \sref{sec:fss:corrections}). It is easy to see that higher order terms in the Taylor expansion of eq.~(\ref{fexbar_taylor}) result in additional corrections $\sim L^{-2k}$, $k > 1$, which are negligible {as well}.

By inserting eq.~(\ref{fexbar_taylor2}) in eq.~(\ref{I_inter}) and neglecting corrections $\sim L^{-2}$ we finally obtain
\begin{equation}
\label{I_casimir}
I\left(\beta, L, L_\parallel\right) = \left[f_{2D}(t) - f_{\rm bulk}(t)\right] + F_C\left(t, L-\frac{1}{2}, L_\parallel\right),
\end{equation}
which still involves the subtraction of the bulk free energy density and the areal free energy density of the two-dimensional layer. Upon substituting eq.~(\ref{casimir_fss}) into the previous expression, we expect the following scaling form for the quantity in eq.~(\ref{I_casimir}):
\begin{eqnarray}
\label{I_casimir_scaling}
\fl I\left(\beta, L, L_\parallel\right) &= \bar{B}(t) + \frac{\theta\left(a_0t(L-1/2+c)^{1/\nu},(L-1/2+c)/L_\parallel\right)}{(L-1/2+c)^3}\nonumber\\
\fl&=\bar{B}(t) + \frac{\theta\left(t\left((L-1/2+c)/\xi_0^+\right)^{1/\nu},\rho\left(1+(c-1/2)/L\right)\right)}{(L-1/2+c)^3}
\end{eqnarray}
where $\bar{B}(t)=f_{2D}(t) - f_{\rm bulk}(t)$ is a $L$-independent background term, and $\theta(\tau,\rho)$ is the universal scaling function associated with the critical Casimir force. It is important to notice that simulations of a lattice with aspect ratio $\rho=L/L_\parallel$ provide data for the scaling variable corresponding to a different aspect ratio
\begin{equation}
\label{mod_rho}
\tilde{\rho}(L)=\rho\left(1+\frac{c-1/2}{L}\right),
\end{equation}
which of course converges to $\rho$ for $L\rightarrow\infty$.

\begin{figure}
\begin{center}
\includegraphics[width=12cm,keepaspectratio]{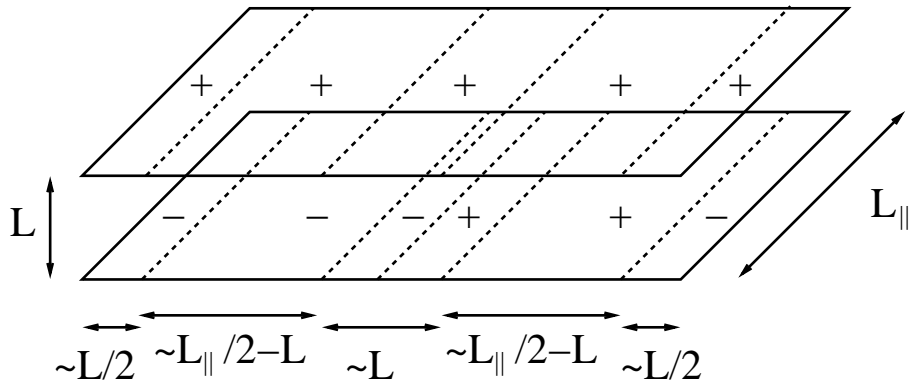}
\end{center}
\caption{Decomposition of a system with a chemical step at the lower surface in the limit of vanishing aspect ratio $\rho=L/L_\parallel\rightarrow 0$. The symbol $+$ ($-$) indicates regions in which the boundary spins are fixed to $+1$ ($-1$) (see the main text). Periodic b.c. are imposed on the lateral directions.}
\label{fig_decomposition}
\end{figure}

In the following, we consider the critical Casimir force for various aspect ratios $\rho$, in particular also the extrapolation to the slab limit, i.e., for $\rho\rightarrow 0$. In this limit the critical Casimir force reduces to the mean value of the critical Casimir force in the $++$ configuration, for which all boundary spins are fixed to the same value, and of the critical Casimir force in the $+-$ configuration, for which at one surface all spins are fixed to $+1$ and at the other surface all are fixed to $-1$.
This can be understood by the following argument.
Near the bulk critical point $T_c$ the actual correlation length in a slab is bounded by $L$ (it may diverge laterally at the film critical point $T_{c,f}(L)<T_c$). Therefore near $T_c$, in the presence of a pair of individual chemical steps, one can divide the system into four pieces: two pieces around each chemical step, which are influenced by the steps, and two lateral pieces which, sufficiently far from the steps, are in the $++$ and $+-$ configuration, respectively (see \fref{fig_decomposition} for an illustration). The former two pieces have a lateral size $\sim L$, while the latter ones have a size $\sim L_\parallel/2-L$. Thus the total critical Casimir force per area is given by
\begin{equation}
\label{force_decomposition}
F_C\simeq\frac{1}{L_\parallel^2}\left[F_{++}L_\parallel(L_\parallel/2-L) + F_{+-}L_\parallel(L_\parallel/2-L) +2F_{p}L_\parallel L\right],
\end{equation}
where, in accordance with \fref{fig_decomposition}, $F_{++}$ is the critical Casimir force per area in the $++$ configuration, $F_{+-}$ is the corresponding one in the $+-$ configuration, and $F_{p}$ is the force per area due to central and lateral pieces.
By taking the limit $L_\parallel\rightarrow\infty$ at fixed $L$, or equivalently, the slab limit $\rho\rightarrow 0$ at fixed $L$, one finds $F_C=(F_{++}+F_{+-})/2$ \footnote{We thank Ettore Vicari for pointing out this argument.}. Along the same line of reasoning one can consider a system in which {\it both} confining walls exhibit a chemical step. In the case that equal (opposite) b.c. face each other, for $\rho\rightarrow 0$ the critical Casimir force is expected to be the same as for laterally homogeneous $++$ ($+-$) b.c. on both sides. This holds even if the two chemical steps are shifted relative to each other. In the case of equal b.c. facing each other and in the presence of a {\it bulk} magnetic field the force again equals the mean value of the ones for the two halves.

\section{Critical Casimir amplitude at $T_c$}
\label{sec:critical}
As discussed in the preceding section, the quantity $I\left(\beta, L, L_\parallel\right)$ can be computed by standard Monte Carlo simulation techniques combined with a numerical integration. At the bulk critical temperature eq.~(\ref{I_casimir_scaling}) becomes
\begin{equation}
\label{I_critical}
I\left(\beta_c, L, L_\parallel\right) = \bar{B} + \frac{\Theta\left(\tilde{\rho}(L)\right)}{(\overline{L}+c)^3},
\end{equation}
where $\bar{B}\equiv \bar{B}(t=0)$ is the non-universal background, $\Theta\left(\rho\right)\equiv\theta\left(0,\rho\right)$ is the universal amplitude of the critical Casimir force at $T_c$, $\overline{L}=L-1/2$, and $\tilde{\rho}(L)$ is given by eq.~(\ref{mod_rho}).

\begin{figure}
\vspace{3em}
\begin{center}
\includegraphics[width=13cm,keepaspectratio]{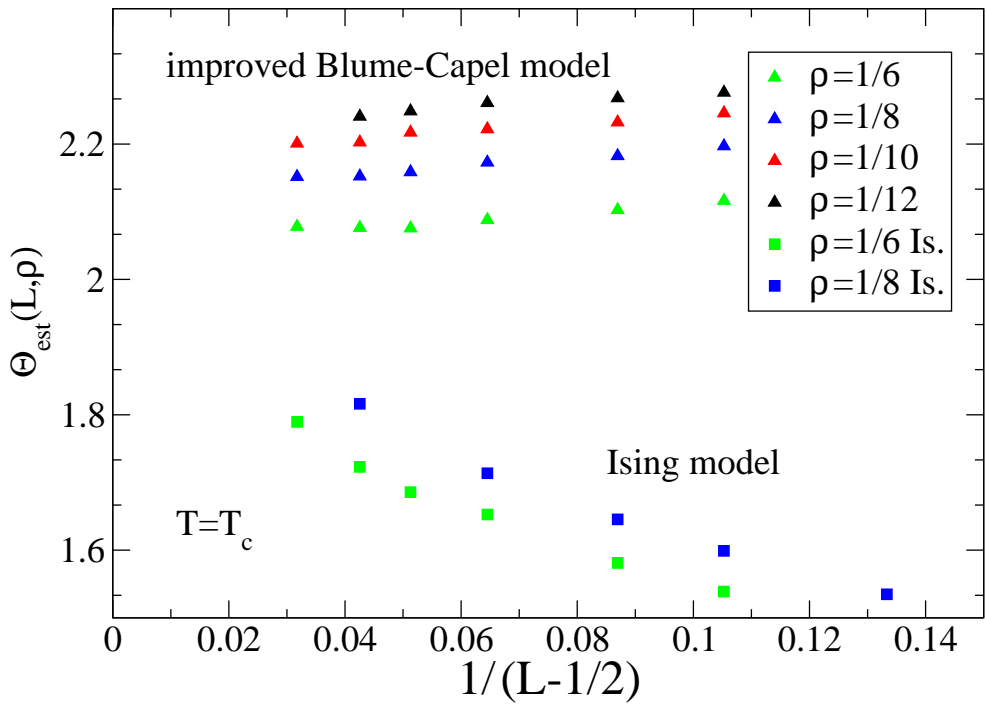}
\end{center}
\caption{The estimator $\Theta_{\rm est}(L, \rho)$ (eq.~(\protect\ref{theta_est})) of the critical Casimir amplitude at $T_c$, for aspect ratios $\rho=1/6$, $1/8$, $1/10$, $1/12$. The simulation data correspond to the system shown in \protect\fref{lattice} for the improved Blume-Capel model (\protect\ref{BC}) and for the standard Ising model (Is.). The statistical error bars of the Monte Carlo data are as large as the symbol size or smaller.}
\label{casimir_amplitude}
\end{figure}

In a series of Monte Carlo simulations, we have evaluated the quantity $I\left(\beta_c, L, L_\parallel\right)$ for lattice sizes $L=8$, $10$, $12$, $16$, $20$, $24$, $32$, $40$, $48$, $64$ and aspect ratios $\rho=1/6$, $1/8$, $1/10$, $1/12$. Certain details of the simulations are reported in \ref{sec:mc}.
From eq.~(\ref{I_critical}) we can obtain an estimator for the critical amplitude $\Theta(\rho)$ by considering the difference concerning two different lattices. To this end we introduce the quantity
\begin{equation}
\label{theta_est}
\Theta_{\rm est}(L, \rho) \equiv \left(L-1/2\right)^3\frac{I\left(\beta_c, L, L_\parallel\right) - I\left(\beta_c, \alpha L, \alpha L_\parallel\right)}{1-\left((L-1/2)/(\alpha L-1/2)\right)^3},
\end{equation}
with a fixed integer $\alpha$. Using eq.~(\ref{I_critical}), one finds that $\Theta_{\rm est}(L\rightarrow\infty, \rho)\rightarrow \Theta(\rho)$, with corrections $\propto 1/\overline{L}$. Moreover such corrections are proportional to $c$ and to $\rho$, i.e., in the slab limit $\rho\rightarrow 0$ and in absence of the leading scaling corrections ($c=0$) the estimator given in eq.~(\ref{theta_est}) equals $\Theta(0)$ up to corrections $\propto L^{-\omega_2}$, $\omega_2=1.67(11)$ \cite{NR-84}. We show the quantity $\Theta_{\rm est}(L, \rho)$ in \fref{casimir_amplitude}, for $\alpha=2$, as obtained by a simulation of the improved Blume-Capel model (\ref{BC}) with $D=0.641$. As a comparison, we also display some data obtained for the standard Ising model corresponding to the Blume-Capel model in the limit $D\rightarrow -\infty$.

The prediction that for the improved model the leading corrections are proportional to $1/\overline{L}$  is confirmed by the data shown in \fref{casimir_amplitude}. Moreover the scaling corrections appear to be numerically small: the data at $L=8$ differ from those at $L=32$ by $3\%$, and the amplitudes for $L\ge 20$, i.e., $1/\overline{L}\lesssim 0.0513$ are all compatible within error bars. On the other hand, the data for the Ising model exhibit stronger corrections; for this model, we expect additional corrections $\propto L^{-\omega}$, with $\omega=0.832(6)$ \cite{Hasenbusch-10}.

In order to obtain an estimate of the critical Casimir amplitude $\Theta(\rho)$, we expand eq.~(\ref{I_critical}) in terms of $1/\overline{L}$ at a fixed aspect ratio $\rho$. In lowest order in $1/\overline{L}$ one obtains
\begin{equation}
\label{I_critical2}
I\left(\beta_c, L, L_\parallel\right) = \bar{B} + \frac{\Theta\left(\rho\right)}{\overline{L}^3} + \frac{C\left(\rho,c\right)}{\overline{L}^4},\qquad \overline{L}\equiv L-1/2,
\end{equation}
where $C\left(\rho,c\right)=-3c\Theta(\rho)+(c-1/2)\rho\Theta'(\rho)$ is the leading correction-to-scaling term, which depends on $\rho$ and $c$. (See also the discussion below leading to eq.~(\ref{C_expansion}).) However eq.~(\ref{I_critical2}) provides an unbiased determination of the critical Casimir amplitude $\Theta\left(\rho\right)$, which is correct even if the corrections to scaling $\propto 1/L$ do not have an analytic origin. We directly fit our Monte Carlo (MC) data for the quantity $I\left(\beta_c=0.3856717, L, L_\parallel\right)$ to eq.~(\ref{I_critical2}) leaving $\bar{B}$, $\Theta$, and $C$ as free parameters. In order to control a possible systematic error due to subleading scaling corrections, we repeat the fit disregarding the smallest lattices. For the various aspect ratios in \tref{fit_critical} we report the fit results as a function of the smallest lattice size $L_{\rm min}$ taken into account for the fit.

\fulltable{\label{fit_critical}Fit of Monte Carlo data to eq.~(\protect\ref{I_critical2}) with free parameters $\bar{B}$, $\Theta$, and $C$. $L_{\rm min}$ is the smallest lattice size taken into account for the fit. $DOF$ denotes degrees of freedom.}
\noindent
\begin{tabular}{@{}l@{\hspace{0.55em}}l@{\hspace{0.55em}}l@{\hspace{0.55em}}l@{\hspace{0.55em}}l}
\br
$L_{\rm min}$  & $\rho=1/6$              & $\rho=1/8$        & $\rho=1/10$              & $\rho=1/12$   \\
\mr
$8$           & $\chi^2/DOF=6.5/7$      & $\chi^2/DOF=14.1/7$ & $\chi^2/DOF=12/7$       & $\chi^2/DOF=3.9/6$\\
              & $\bar{B}=0.03235167(9)$ & $\bar{B}=0.03235196(8)$  & $\bar{B}=0.03235157(10)$ & $\bar{B}=0.0323514(1)$\\
              & $\Theta=2.048(3)$       & $\Theta=2.132(2)$  & $\Theta=2.190(3)$        & $\Theta=2.229(3)$\\
              & $C=0.60(2)$             & $C=0.55(2)$        & $C=0.47(3)$              & $C=0.41(2)$\\
\mr
$10$          & $\chi^2/DOF=6.1/6$      & $\chi^2/DOF=7.6/6$ & $\chi^2/DOF=7.3/6$      & $\chi^2/DOF=2.9/5$\\
              & $\bar{B}=0.0323517(1)$  & $\bar{B}=0.03235210(9)$  & $\bar{B}=0.0323517(1)$  & $\bar{B}=0.0323515(2)$\\
              & $\Theta=2.046(4)$       & $\Theta=2.124(3)$  & $\Theta=2.181(4)$       & $\Theta=2.225(4)$\\
              & $C=0.63(4)$             & $C=0.65(3)$        & $C=0.57(4)$             & $C=0.46(4)$\\
\mr
$12$          & $\chi^2/DOF=5.3/5$      & $\chi^2/DOF=7.2/5$  & $\chi^2/DOF=6.9/5$     & $\chi^2/DOF=2.5/4$\\
              & $\bar{B}=0.0323517(1)$  & $\bar{B}=0.03235206(10)$ & $\bar{B}=0.0323517(1)$  & $\bar{B}=0.0323516(2)$\\
              & $\Theta=2.050(6)$       & $\Theta=2.126(4)$  & $\Theta=2.184(5)$       & $\Theta=2.221(6)$\\
              & $C=0.56(7)$             & $C=0.61(5)$        & $C=0.52(7)$             & $C=0.51(7)$\\
\mr
$16$          & $\chi^2/DOF=5.3/4$      & $\chi^2/DOF=4.9/4$ & $\chi^2/DOF=6.6/4$      & $\chi^2/DOF=0.6/3$\\
              & $\bar{B}=0.0323517(2)$  & $\bar{B}=0.0323522(1)$   & $\bar{B}=0.0323517(2)$  & $\bar{B}=0.0323518(2)$\\
              & $\Theta=2.05(1)$        & $\Theta=2.112(8)$  & $\Theta=2.18(1)$        & $\Theta=2.20(1)$\\
              & $C=0.5(2)$              & $C=0.9(2)$         & $C=0.6(2)$              & $C=0.8(2)$\\
\br
\end{tabular}
\endfulltable

Inspecting the fit results, we generally observe a good $\chi^2/DOF$ ($DOF$ is the number of degrees of freedom, i.e., the number of statistically independent points minus the number of fit parameters) for $L_{\rm min} \ge 10$ and the results appear to be stable with respect to $L_{\rm min}$. While there is a clear dependence of the Casimir amplitude $\Theta$ on $\rho$, as expected (see eq.~(\ref{I_casimir_scaling})) the background term $\bar{B}$ does not exhibit a dependence on $\rho$. We observe that $\bar{B}$ as determined at $\rho=1/8$ is slightly shifted with respect to the corresponding values determined for the other aspect ratios. However the difference is tiny (two error bars in the worst case). Thus we conclude that the observed shift can be interpreted as a statistical fluctuation; note also that the ratio $\chi^2/DOF$ is slightly worse for the data at $\rho=1/8$.

Corrections to eq.~(\ref{I_critical2}) are generated by next-to-leading irrelevant operators and result in an additional term $\propto L^{-3-\omega_2}$, with $\omega_2=1.67(11)$ \cite{NR-84}. Fits including such correction do not result in significant deviations from the results given in \tref{fit_critical}.

By judging conservatively the variation of the resulting $\Theta$ with respect to $L_{\rm min}$, from \tref{fit_critical} we obtain the following estimates:
\begin{eqnarray}
\label{Delta6}
&\Theta(1/6)=2.048(6),\\
\label{Delta8}
&\Theta(1/8)=2.126(5),\\
\label{Delta10}
&\Theta(1/10)=2.183(6),\\
\label{Delta12}
&\Theta(1/12)=2.223(7).
\end{eqnarray}

\begin{figure}
\vspace{3em}
\begin{center}
\includegraphics[width=12cm,keepaspectratio]{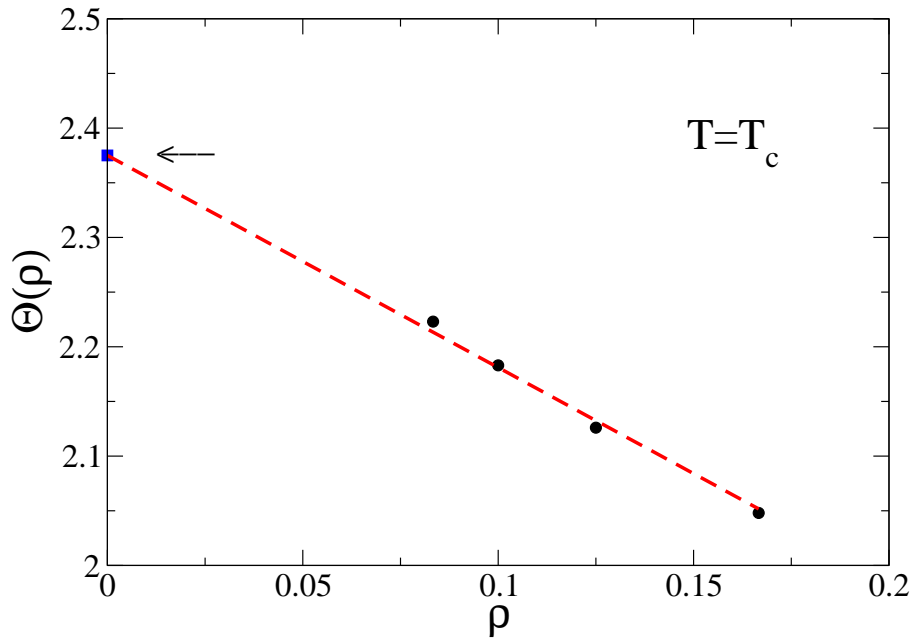}
\end{center}
\caption{Critical Casimir force amplitude $\Theta(\rho)=\theta(0,\rho)$ (see eqs.~(\protect\ref{casimir_fss_leading}) and (\protect\ref{theta_Delta})) at $T_c$ for aspect ratios $\rho=1/6$, $1/8$, $1/10$, $1/12$, as inferred from \protect\tref{fit_critical} (eqs.~(\protect\ref{Delta6})-(\protect\ref{Delta12})), as well as the limit $\rho\rightarrow 0$ reported in eq.~(\ref{Theta}). The dashed line represents eq.~(\protect\ref{theta_expansion}), with its parameters given by eqs.~(\protect\ref{Theta}) and (\protect\ref{D}). As expected on the basis of eq.~(\protect\ref{taylor}), $\Theta(\rho\rightarrow 0)$ does not exhibit a quadratic term. The statistical error bars have the same size as the symbols.}
\label{critvsrho}
\end{figure}

These amplitudes are shown in \fref{critvsrho}. They clearly show a linear dependence on the aspect ratio $\rho$. This allows us to determine the limit $\Theta(\rho\rightarrow 0)$ by expanding
\begin{equation}
\label{theta_expansion}
\Theta(\rho)=\Theta(0) + E\rho + O(\rho^3),
\end{equation}
neglecting terms which are nonlinear in $\rho$. As discussed at the end of \sref{sec:fss:energies}, in the expansion of eq.~(\ref{theta_expansion}), the amplitude of the critical Casimir force at $T_c$ has no quadratic term $\propto \rho^2$. Thus corrections to the above expression are expected to be at least of third order, i.e., $\propto\rho^3$. By inserting eq.~(\ref{theta_expansion}) into eq.~(\ref{I_critical}) and using eq.~(\ref{mod_rho}) we obtain in linear order in $\rho$
\begin{equation}
\label{I_critical_rho}
\fl I\left(\beta_c, L, L_\parallel\right) = \bar{B} + \frac{\Theta(0)+E\rho\left(1+(c-1/2)/(\overline{L}+1/2)\right)}{(\overline{L}+c)^3},\qquad \overline{L}\equiv L-1/2.
\end{equation}
Inspection of eq.~(\ref{I_critical_rho}) shows that scaling corrections $\propto 1/L$ for $I$ emerge from two contributions. The first contribution stems from the dependence on the aspect ratio $\rho$: the MC data of a system with sizes $L\times L_\parallel\times L_\parallel$, corresponding to $\rho=L/L_\parallel$, result in the force for a system with perpendicular size $\overline{L}=L-1/2$, which has an aspect ratio $(L-1/2)/L_\parallel=\rho(1-1/(2L))$. This correction is proportional to $\rho$. The second source of scaling corrections is due to the non-periodic boundaries and it is proportional to $c$.

We first consider fits of the MC data ignoring scaling corrections due to the boundary conditions. We fit all data for the various aspect ratios to eq.~(\ref{I_critical_rho}), setting $c=0$ and leaving $\bar{B}$, $\Theta(0)$, and $E$ as free parameters. The results for those fits are reported in \tref{fit_critical_all_noc}.

\Table{\label{fit_critical_all_noc}Fits of Monte Carlo data to eq.~(\protect\ref{I_critical_rho}), with free parameters $\bar{B}$, $\Theta(0)$, $E$ and setting $c=0$. $L_{\rm min}$ is the smallest lattice size taken into account for the fit. $DOF$ denotes the degrees of freedom.}
\begin{tabular}{@{}lllll}
\br
$L_{\rm min}$  & $\bar{B}$       & $\Theta(0)$ & $E$         & $\chi^2/DOF$\\
\mr
$8$           & $0.03235066(4)$ & $2.4366(9)$ & $-2.026(8)$ & $883/36$ \\
$10$          & $0.03235100(4)$ & $2.431(1)$  & $-2.05(1)$  & $342/32$ \\
$12$          & $0.03235128(4)$ & $2.425(2)$  & $-2.07(2)$  & $133/28$ \\
$16$          & $0.03235153(5)$ & $2.415(3)$  & $-2.06(3)$  & $59/24$ \\
\br
\end{tabular}
\endTable

These fits have a large ratio $\chi^2/DOF$ and in addition the fitted parameters show a systematic drift, which is larger than the statistical error. From this we conclude that scaling corrections due to the parameter $c$ in eq.~(\ref{I_critical_rho}) are sizeable within the statistical precision of the MC data. In view of this we fit all data for the various aspect ratios to eq.~(\ref{I_critical_rho}), leaving $\bar{B}$, $\Theta(0)$, $E$, and $c$ as free parameters. The results of these fits are reported in \tref{fit_critical_all}.

\Table{\label{fit_critical_all}Fits of the Monte Carlo data to eq.~(\protect\ref{I_critical_rho}) with free parameters $B$, $\Theta(0)$, $E$, and $c$. $L_{\rm min}$ is the smallest lattice size taken into account in the fit. $DOF$ denotes the degrees of freedom.}
\begin{tabular}{@{}llllll}
\br
$L_{\rm min}$  & $\bar{B}$       & $\Theta(0)$ & $E$         & $c$         & $\chi^2/DOF$\\
\mr
$8$           & $0.03235170(5)$ & $2.386(2)$  & $-1.991(8)$ & $-0.058(2)$ & $56/35$ \\
$10$          & $0.03235181(5)$ & $2.383(2)$  & $-2.02(1)$  & $-0.069(3)$ & $35/31$ \\
$12$          & $0.03235180(6)$ & $2.388(3)$  & $-2.05(2)$  & $-0.066(5)$ & $32/27$ \\
$16$          & $0.03235188(7)$ & $2.377(6)$  & $-2.04(3)$  & $-0.09(1)$  & $27/23$ \\
\br
\end{tabular}
\endTable

Here the situation is much improved and fits with $L\ge 10$ show a good ratio $\chi^2/DOF$. Moreover, the results are stable upon increasing the smallest lattice size taken into account, which underscores the quality of the fits. Accordingly we obtain as final estimates
\begin{eqnarray}
\label{B}
&\bar{B}=0.03235181(7),\\
\label{Theta}
&\Theta(0)=2.386(5),\\
\label{D}
&E=-2.04(3),\\
\label{c}
&c=-0.066(5).
\end{eqnarray}
Also in this case we have performed fits of our Monte Carlo data by adding a next-to-leading scaling correction term to eq.~(\ref{I_critical_rho}). These results do not exhibit significant deviations from eqs.~(\ref{B})-(\ref{c}).
In \fref{critvsrho} we compare the amplitude of the Casimir force at $T_c$ for the various aspect ratios reported in eqs.~(\ref{Delta6})-(\ref{Delta12}) with the linear dependence on $\rho$ predicted by eq.~(\ref{theta_expansion}), using eqs.~(\ref{Theta}) and (\ref{D}). We find very good agreement. 

As discussed at the end of \sref{sec:simulations}, in the limit $\rho\rightarrow 0$ the critical Casimir force reduces to the mean value of the forces valid for the $++$ and the $+-$ configuration, respectively, where the first case corresponds to a system in which all boundary spins are fixed to the same value, and the latter case corresponds to a system in which on one surface the spins are fixed to $+1$ and on the other to $-1$. These laterally homogeneous Ising systems have been investigated in Refs.~\cite{VGMD-07,VGMD-08,Hasenbusch-10c} by Monte Carlo simulations. According to Ref.~\cite{VGMD-08}, $\Theta_{++}=-0.76(6)$ and $\Theta_{+-}=5.42(4)$, so that $(\Theta_{++}+\Theta_{+-})/2=2.33(4)$, in marginal agreement with eq.~(\ref{Theta}). According to Ref.~\cite{Hasenbusch-10c}, $\Theta_{++}=-0.820(15)$ and $\Theta_{+-}=5.613(20)$, so that $(\Theta_{++}+\Theta_{+-})/2=2.396(13)$, in perfect agreement with eq.~(\ref{Theta}). In Refs.~\cite{VGMD-07,VGMD-08} simulations have been carried out for the standard Ising model, for which the scaling corrections proportional to $L^{-0.8}$ and those proportional to $L^{-1}$ are difficult to disentangle. Thus we expect our results to be more reliable with respect to those of Refs.~\cite{VGMD-07,VGMD-08}, because in our model scaling corrections are under control. Moreover, our results are in perfect agreement with those of Ref.~\cite{Hasenbusch-10c}, in which the improved Blume-Capel model has been used.

Expansion of eq.~(\ref{I_critical_rho}) to the lowest order in the scaling corrections yields, up to terms linear in the aspect ratio $\rho$,
\begin{equation}
I\left(\beta_c, L, L_\parallel\right) = \bar{B} + \frac{\Theta(0)+E\rho}{\overline{L}^3}+ \frac{-3c\Theta(0)-2\rho cE -\rho E/2}{\overline{L}^4}.
\end{equation}
Comparing this result with eq.~(\ref{I_critical2}) and taking into account eq.~(\ref{theta_expansion}), we obtain an expression for the amplitude $C(\rho,c)$ which appears in eq.~(\ref{I_critical2}):
\begin{equation}
\label{C_expansion}
C\left(\rho,c\right)=-3c\Theta(0)-2\rho cE-\frac{1}{2}\rho E.
\end{equation}
Thus in line with the expression for $C\left(\rho,c\right)$ given after eq.~(\ref{I_critical2}), the amplitude $C\left(\rho,c\right)$ of the correction to scaling in eq.~(\ref{I_critical2}) can be expressed in terms of $c$ and the critical Casimir amplitude $\Theta(\rho)$. This relationship is due to the analytic origin of the corrections to scaling (see also the discussion at the end of \sref{sec:fss:corrections}). Using the results in eqs.~(\ref{Theta})-(\ref{c}) we obtain $C(1/6)\simeq 0.60(3)$, $C(1/8)\simeq 0.57(4)$, $C(1/10)\simeq 0.55(4)$, and $C(1/12)\simeq 0.53(4)$. Comparing these values with those shown in \tref{fit_critical}, for the most reliable data with $L_{\rm min}=10$ and $L_{\rm min}=12$ we observe agreement within the error bars. Given the limited available precision, this is a non-trivial consistency check of our scaling ansatz.

\section{Casimir scaling function}
\label{sec:theta}
\subsection{General results}
\label{sec:theta:general}
In order to determine the full scaling function $\theta(\tau,\rho)$, in eq.~(\ref{I_casimir_scaling}) we have to subtract the $L-$independent term $\bar{B}(t)$ from the quantity $I(\beta,L,L_\parallel)$ sampled by MC simulations. Moreover, eq.~(\ref{I_casimir_scaling}) tells that the MC data obtained at an aspect ratio $\rho$ refers to the Casimir scaling function at a \emph{modified} aspect ratio $\tilde{\rho}$ given by eq.~(\ref{mod_rho}). Since we have determined the temperature-independent length $c$ at $T_c$ (see eq.~(\ref{c})), we can calculate the quantity $\tilde{\rho}$ for all MC data. In order to subtract the background term $\bar{B}(t)$ we proceed as follows. First, to avoid dealing with the normalization of the scaling variable appearing in eqs.~(\ref{exp_ut}), (\ref{xi_crit}), and (\ref{xivsut}) we shall consider the unnormalized scaling function $\overline{\theta}(x,\rho)$ defined as
\begin{equation}
\label{unnormalized}
\overline{\theta}(tL^{1/\nu},\rho)\equiv\theta(a_0t L^{1/\nu},\rho),
\end{equation}
where in eq.~(\ref{exp_ut}) we have considered only the leading term in the expansion of $u_t$; here, according to eq.~(\ref{xivsut}), $a_0$ is an amplitude not yet specified.

We have already found in \sref{sec:critical} that for $\rho\rightarrow 0$ the amplitude $\Theta$ of the critical Casimir force at $T_c$ exhibits a linear dependence in $\rho$. This leads us to introduce the following corresponding generalizations:
\begin{eqnarray}
\label{theta_rho}
&\theta(\tau,\rho)=\theta(\tau,0) + \rho E(\tau) + O(\rho^2), \\
\label{theta_rho_unnormalized}
&\overline{\theta}(\bar{\tau},\rho)=\overline{\theta}(\bar{\tau},0) + \rho\overline{E}(\bar{\tau}) + O(\rho^2),
\end{eqnarray}
omitting possible higher-order terms in the preceding expansion.
For each given pair $(\beta,L)$, inserting eq.~(\ref{theta_rho}) or eq.~(\ref{theta_rho_unnormalized}) into eq.~(\ref{I_casimir_scaling}) leads to a linear dependence on $\rho$ of the quantity $I(\beta,L,L_\parallel)$. Accordingly, a simple linear interpolation scheme for the MC data from systems with different aspect ratios $\rho$ yields, for every given pair $(\beta,L)$,
\begin{equation}
\label{I_tilde}
\tilde{I}\left(\beta,L,\rho\right) = \bar{B}(t) + \frac{\overline{\theta}\left(t(L-1/2+c)^{1/\nu},\rho\right)}{(L-1/2+c)^3},
\end{equation}
with $c$ from eq.~(\ref{c}). There are two reasons for interpolating the data at aspect ratio $\rho$ instead of directly using the data at $\tilde{\rho}$ as it was done in \sref{sec:critical}. First, in this way we have suppressed the $1/L$ correction which arises from eq.~(\ref{mod_rho}). Secondly, as it will become clear in the following, by considering data for various lattice sizes with the {\it same} aspect ratio $\rho$, we are left with a dependence on $L$ only in the first argument of the scaling function $\theta(\tau,\rho)$, a fact which enables us to eliminate the correction to scaling in an easy way.

Next, we define a function $g$ by taking the difference, at the same aspect ratio $\rho$, of the two expressions for $\tilde{I}$ corresponding to lattice sizes $L$ and $\alpha L$, respectively:
\begin{equation}
\label{gdef}
g(\beta,L,\rho)\equiv \left(L-1/2+c\right)^3\left[\tilde{I}\left(\beta,L,\rho\right)-\tilde{I}\left(\beta,\alpha L,\rho\right)\right].
\end{equation}
As in eq.~(\ref{theta_est}) for \fref{casimir_amplitude}, we choose $\alpha=2$. Inserting eq.~(\ref{I_tilde}) into eq.~(\ref{gdef}) leads to
\begin{eqnarray}
\label{gscaling}
\fl g(\beta,L,\rho) = \overline{\theta}\left(t(L-1/2+c)^{1/\nu},\rho\right) - A(\alpha, L)\overline{\theta}\left(B(\alpha,L)t(L-1/2+c)^{1/\nu},\rho\right),\nonumber\\
\fl A(\alpha,L) \equiv \left(\frac{L-1/2+c}{\alpha L-1/2+c}\right)^3,\qquad B(\alpha,L) \equiv \left(\frac{\alpha L-1/2+c}{L-1/2+c}\right)^{1/\nu}.
\end{eqnarray}
This eliminates the background term $\bar{B}(t)$. With a slight abuse of notation on the left hand side, we can rewrite the previous equation as
\begin{equation}
\label{gscaling2}
\fl g(\bar{\tau},\rho; L,\alpha) = \overline{\theta}\left(\bar{\tau},\rho\right) - A(\alpha, L)\overline{\theta}\left(B(\alpha,L)\bar{\tau},\rho\right),\qquad \bar{\tau}=t(L-1/2+c)^{1/\nu}.
\end{equation}
If we had simply used $I$ instead of $\tilde{I}$, the two terms in eq.~(\ref{gscaling2}) would refer to the scaling function $\theta$ calculated at different aspect ratios, leaving us with a complicated expression. With our choice $\alpha=2$, we have $A(\alpha,L)=1/8 + O(1/L)$ and $B(\alpha,L)=2^{1/\nu} + O(1/L)\simeq 3 + O(1/L)$. Keeping in mind that for large $\tau$ the Casimir force decays exponentially, we see that in eq.~(\ref{gscaling2}) the second term represents a correction to the first term, which vanishes in the limit $\alpha\rightarrow\infty$. In order to eliminate this correction we introduce
\begin{eqnarray}
\label{gn}
g_n(\bar{\tau},\rho; L,\alpha) &\equiv \sum_{k=0}^n\left[A(\alpha,L)\right]^kg(\left[B(\alpha,L)\right]^k\bar{\tau},\rho; L,\alpha),\nonumber\\
g_0(\bar{\tau},\rho; L,\alpha) &= g(\bar{\tau},\rho; L,\alpha).
\end{eqnarray}
Inserting eq.~(\ref{gscaling2}) into eq.~(\ref{gn}) we obtain
\begin{equation}
\label{gnscaling}
g_n(\bar{\tau},\rho; L,\alpha) = \overline{\theta}\left(\bar{\tau},\rho\right) - \left[A(\alpha, L)\right]^{n+1}\overline{\theta}\left(\left[B(\alpha,L)\right]^{n+1}\bar{\tau},\rho\right).
\end{equation}
Thus we have $g_n(\bar{\tau},\rho;L,\alpha)\ \vector(1,0){25}\hspace{-2em}^{n\rightarrow\infty}\ \overline{\theta}\left(\bar{\tau},\rho\right)$. Moreover, the error due to truncating the sum in eq.~(\ref{gn}) is proportional to $\left[A(\alpha=2, L)\right]^{n+1}\simeq 2^{-3(n+1)}$. Accordingly, the sum in eq.~(\ref{gn}) converges quickly to the scaling function $\overline{\theta}$. Starting from a Monte Carlo estimate of $g$ at a certain value of $\bar{\tau}$, in order to be able to calculate the sum appearing in eq.~(\ref{gn}) the value of the function $g$ at $\bar{\tau}'=B^k\bar{\tau}$ is required, which might be not directly available from the MC data. However, this value can be estimated by using a simple interpolation spline for those values of the function $g$ which are available. With the present precision of our data, at $n=2$ the error associated with the truncation and given by eq.~(\ref{gnscaling}) is smaller than the statistical error bars. Therefore we use the approximation $\overline{\theta}(\bar{\tau},\rho)\simeq g_2(\bar{\tau},\rho;L,\alpha)$. We note that, if the procedure is correct, the reconstructed function $\overline{\theta}$ should not depend explicitly on $\alpha$ and $L$, but only on the scaling variable $\bar{\tau}$ and the aspect ratio $\rho$. Finally, we implement the appropriate normalization by using eq.~(\ref{unnormalized}) with the normalization constant $a_0$ given in eqs.~(\ref{xivsut}) and (\ref{xi0}).

\begin{figure}
\vspace{3em}
\begin{center}
\includegraphics[width=12cm,keepaspectratio]{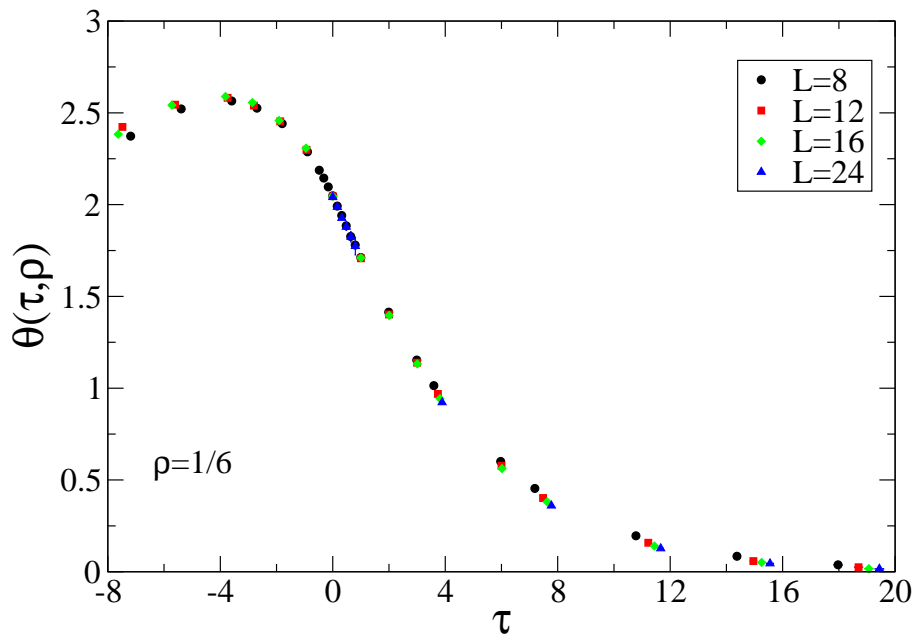}
\end{center}
\caption{The universal scaling function $\theta(\tau,\rho)$ of the critical Casimir force (eq.~(\protect\ref{casimir_fss_leading})) for $\tau=t(L/\xi_0^+)^{1/\nu}$ and for an aspect ratio $\rho=1/6$. The error bars are smaller than the symbol sizes.}
\label{thetar6}
\end{figure}

\begin{figure}
\begin{center}
\includegraphics[width=12cm,keepaspectratio]{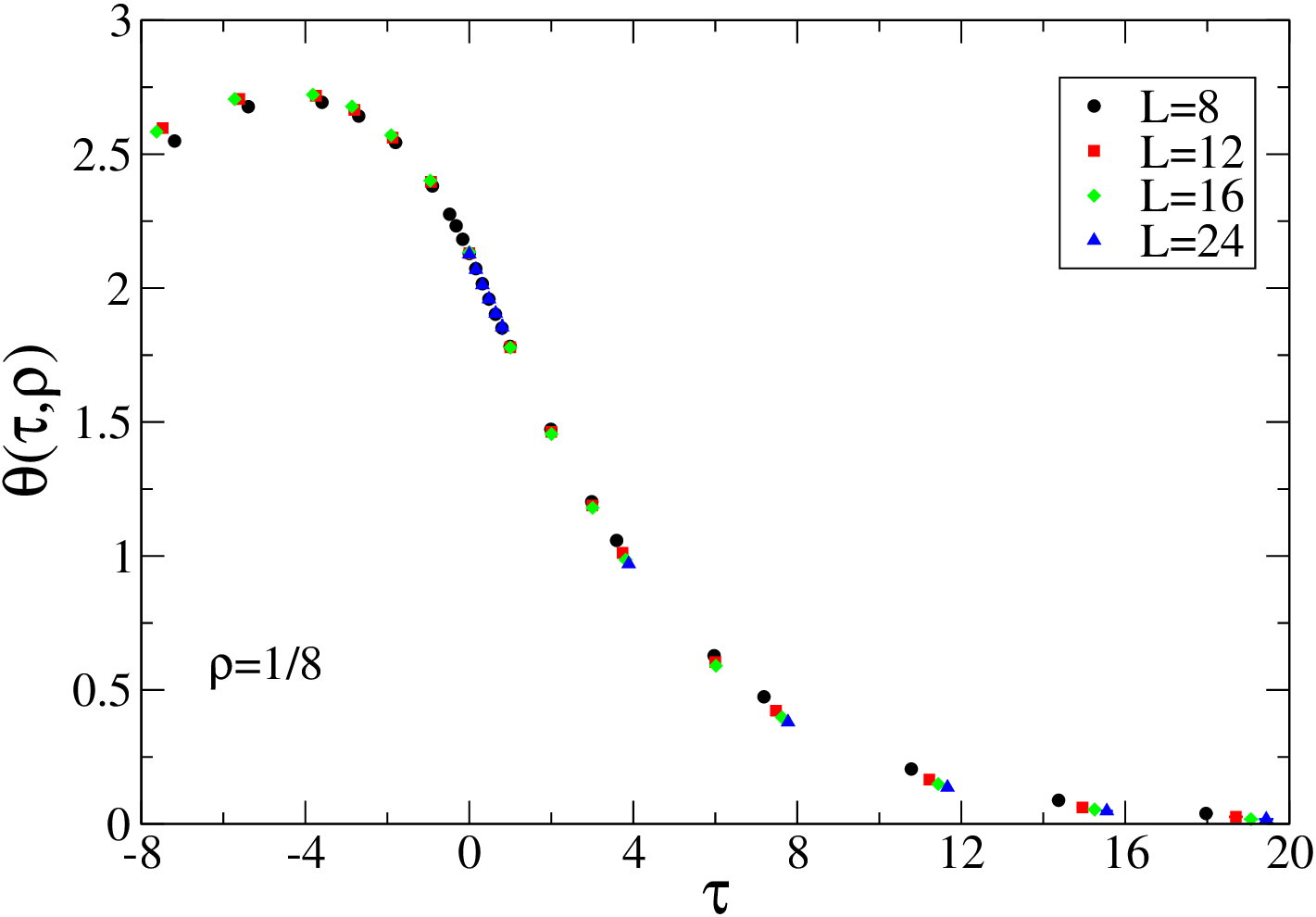}
\end{center}
\caption{Same as \protect\fref{thetar6} for the aspect ratio $\rho=1/8$.}
\label{thetar8}
\end{figure}

\begin{figure}
\vspace{3em}
\begin{center}
\includegraphics[width=12cm,keepaspectratio]{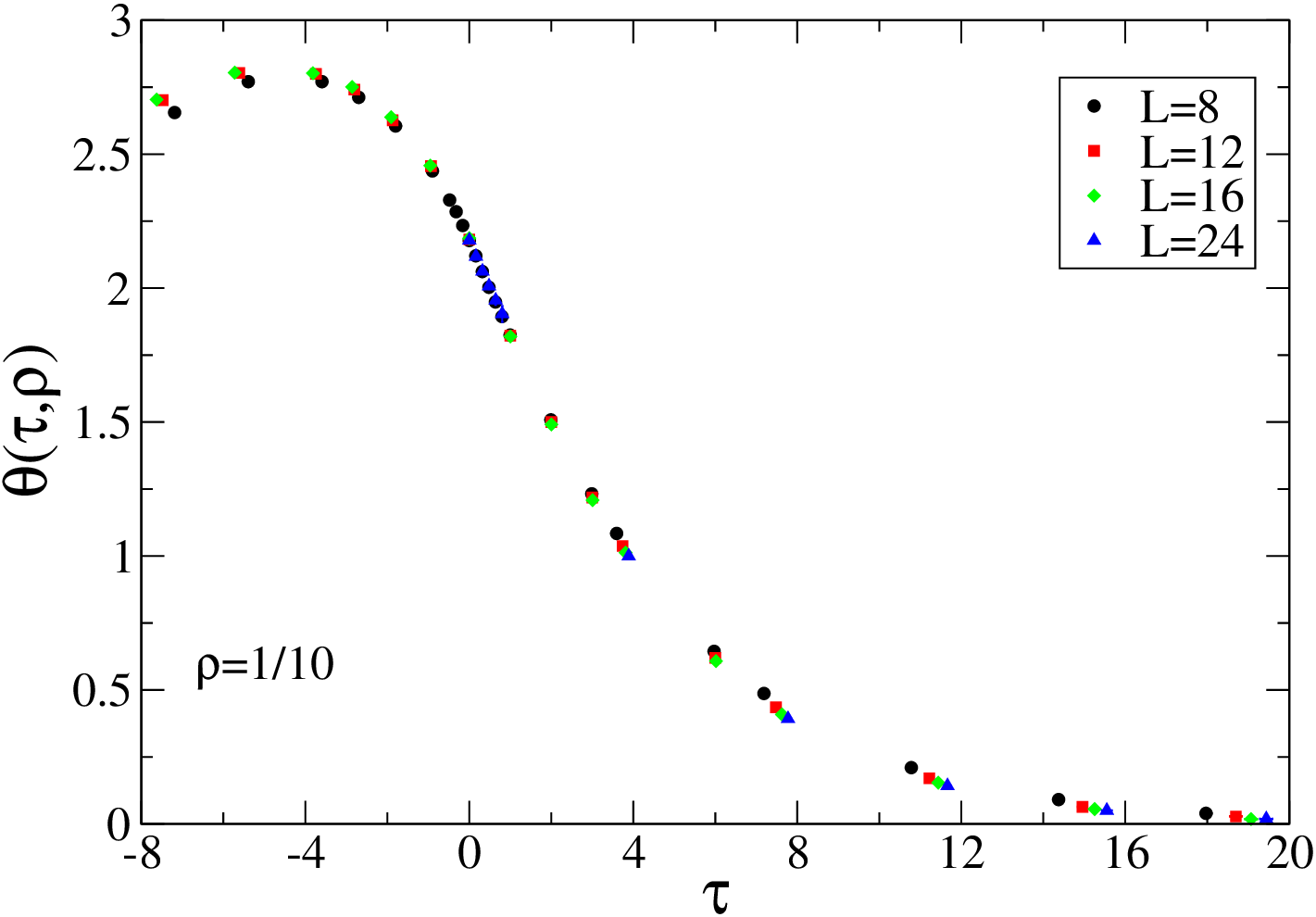}
\end{center}
\caption{Same as \protect\fref{thetar6} for the aspect ratio $\rho=1/10$.}
\label{thetar10}
\end{figure}

\begin{figure}
\begin{center}
\includegraphics[width=12cm,keepaspectratio]{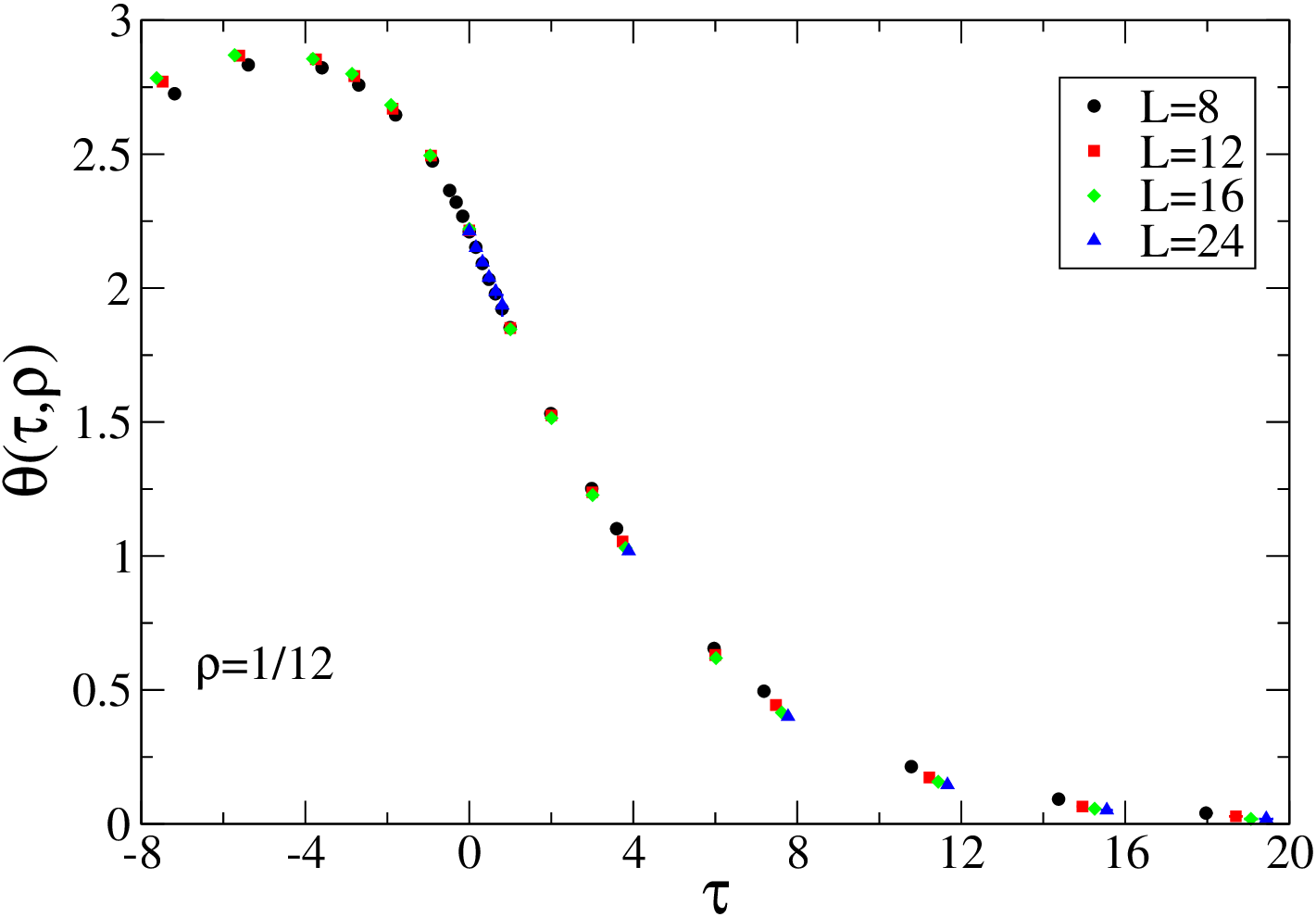}
\end{center}
\caption{Same as \protect\fref{thetar6} for the aspect ratio $\rho=1/12$.}
\label{thetar12}
\end{figure}

In figures~\ref{thetar6}, \ref{thetar8}, \ref{thetar10}, and \ref{thetar12} we show the universal scaling function $\theta(\tau,\rho)$ for aspect ratios $\rho=1/6$, $1/8$, $1/10$, and $1/12$, respectively. In order to treat correctly the statistical covariance between the various quantities in eq.~(\ref{gn}), the error bars have been calculated according to the jackknife procedure (see, e.g., Ref.~\cite{AM-book}). We observe a good data collapse for $L\ge 12$, which supports qualitatively our procedure and the scaling ansatz in eq.~(\ref{I_casimir_scaling}). The data for $L=8$ appear to be slightly off the curves obtained for larger lattices. This is not surprising because, as we already noted in \sref{sec:critical}, data for such a small lattice size suffer from higher-order scaling corrections.

\begin{figure}
\vspace{3em}
\begin{center}
\includegraphics[width=13cm,keepaspectratio]{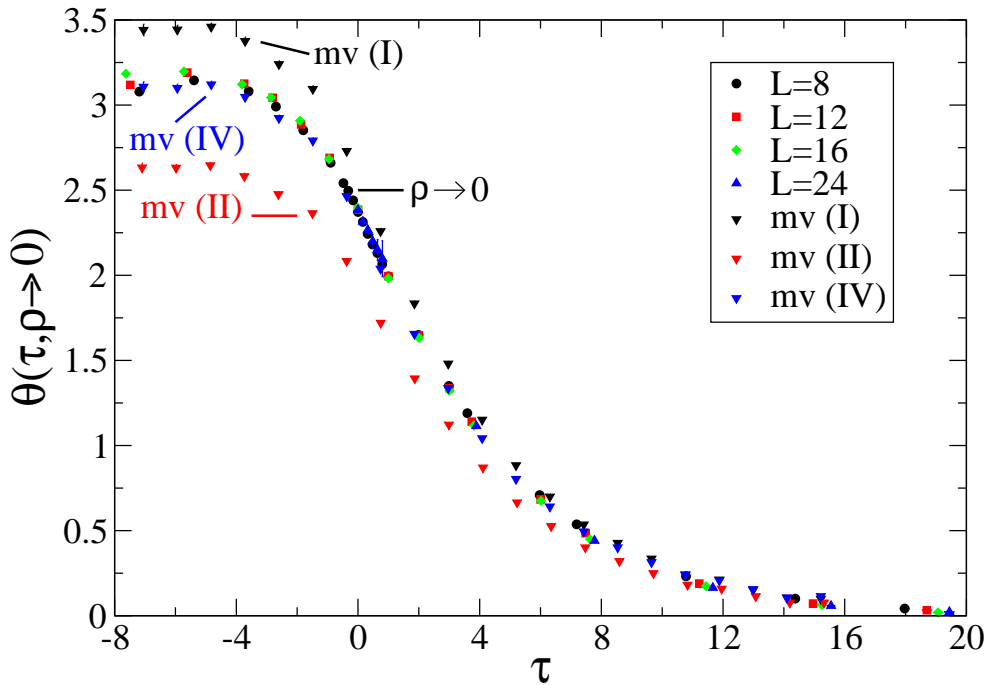}
\end{center}
\caption{The universal scaling function $\theta(x,\rho)$ of the critical Casimir force (eq.~(\protect\ref{casimir_fss_leading})) as a function of the scaling variable $\tau=t(L/\xi_0^+)^{1/\nu}$ and extrapolated to the aspect ratio $\rho=0$. We compare our results with the mean value (mv) of the critical Casimir forces for laterally homogeneous $++$ and $+-$ b.c., as obtained from the MC data in Ref.~\protect\cite{VGMD-08} for the three approximants I, II, and IV presented therein. There is satisfactory agreement with approximant IV, which was claimed to be the best one among those studied in Ref.~\protect\cite{VGMD-08}.}
\label{thetar0}
\end{figure}

In \fref{thetar0} we present the scaling function for the critical Casimir force extrapolated to $\rho\rightarrow 0$. As we mentioned at the end of \sref{sec:simulations}, in this limit the Casimir force is expected to be the mean value of the force for the laterally homogeneous $++$ and $+-$ configurations. The scaling functions for these b.c. have been computed by Monte Carlo simulations in Ref.~\cite{VGMD-08}, where $3$ curves, denoted as I, II, and IV, have been presented. From those curves we have formed their mean values by using an interpolation spline for the three approximants presented in Ref.~\cite{VGMD-08}. In \fref{thetar0} we show a comparison with our results. There is good agreement between our curve and the one obtained from the approximant IV of Ref.~\cite{VGMD-08}. Incidentally, it is reported in Ref.~\cite{VGMD-08} that among the three it is this approximant which describes the finite-size scaling of the MC data best.

\begin{figure}
\vspace{3em}
\begin{center}
\includegraphics[width=13cm,keepaspectratio]{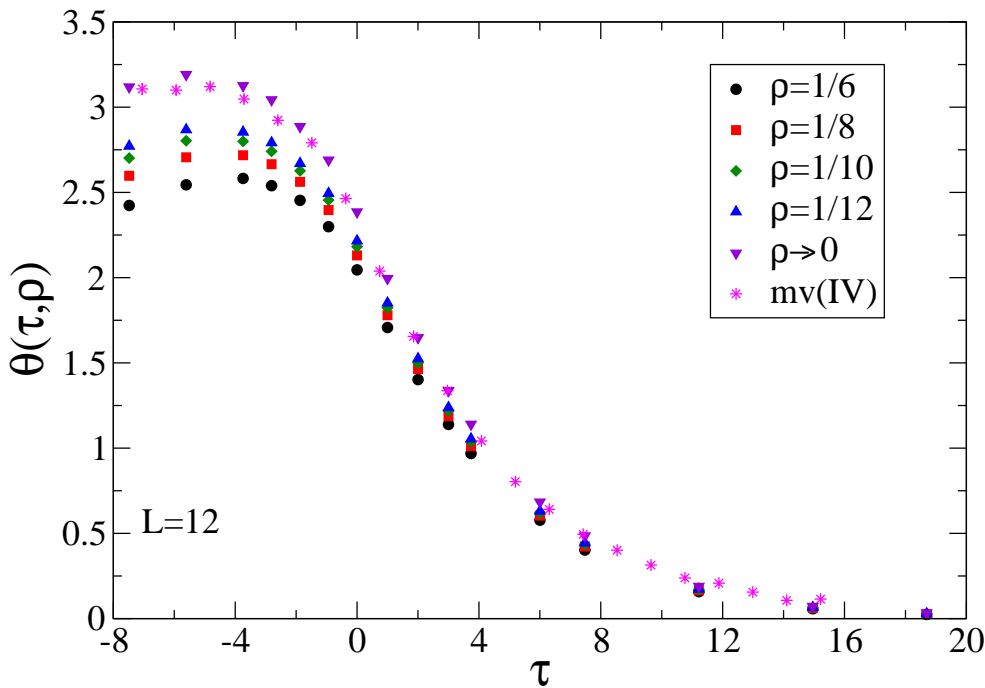}
\end{center}
\caption{Comparison of the universal scaling function $\theta(x,\rho)$ of the critical Casimir force (eq.~(\protect\ref{casimir_fss_leading})) as a function of the scaling variable $\tau=t(L/\xi_0^+)^{1/\nu}$ for the aspect ratios $\rho=1/6$, $1/8$, $1/10$, $1/12$, and $\rho\rightarrow 0$ for fixed $L=12$.
We also show a comparison of our results with the mean value (mv) of the critical Casimir forces for laterally homogeneous $++$ and $+-$ b.c., as obtained from the MC data in Ref.~\protect\cite{VGMD-08} for the approximant IV presented therein.}
\label{thetacomparison}
\end{figure}

In \fref{thetacomparison} we show a comparison of the scaling function of the critical Casimir force for the aspect ratios $\rho=1/6$, $1/8$, $1/10$, $1/12$, $0$, as obtained for $L=12$, as well as the mean value of the force for the laterally homogeneous $++$ and $+-$ b.c. following from approximant IV of Ref.~\cite{VGMD-08}. Within the available range of $\tau$ values, the function $\theta(\tau,\rho)$ increases for decreasing $\rho$.

\subsection{Chemical-steps contribution}
\label{sec:theta:step}
As discussed at the end of \sref{sec:simulations}, in the limit $\rho\rightarrow 0$ the critical Casimir force in the presence of a pair of individual chemical steps on one of the confining walls (\fref{lattice}) reduces to the mean value of the forces for the laterally homogeneous $++$ and $+-$ configurations. This mean value can be interpreted as the force for an ersatz system in which the system shown in \fref{lattice} is decomposed into two halves disconnected along the chemical steps which become infinitely separated in the limit $\rho\rightarrow 0$. Accordingly one expects that the effect of the presence of the pair of individual chemical steps on the critical Casimir force enters into its dependence on the aspect ratio $\rho$. This consideration can be formalized by generalizing the discussion presented at the end of \sref{sec:fss:energies}. In the following we first discuss how to define a line contribution to the singular part of the free energy density for general b.c., specializing later the argument to the present case. To this end we consider two slabs $L\times L_\parallel\times L_\parallel$, which undergo a second-order phase transition, with the same bulk and surface universality classes. The confining surfaces of the first slab $(\mathit{1})$ exhibit laterally homogeneous b.c., such that the system displays two surfaces but no edges. This can be realized by imposing periodic b.c. in both lateral directions and homogeneous b.c. on the two confining surfaces. The second slab $(\mathit{2})$ has one or more edges of extension $L_\parallel$ or has {a pair of individual inhomogeneities} of linear extension $L_\parallel$. The first case can be realized by imposing periodic b.c. only in one of the two lateral directions corresponding to the linear size $L_\parallel$, leaving open b.c. in the other directions: such a slab has two surfaces of area $L_\parallel\times L_\parallel$ and four edges of extension $L_\parallel$. The second case can be realized by imposing a laterally inhomogeneous b.c. with linear spatial extent $L_\parallel$, as in the system shown in \fref{lattice}. For these systems $(\mathit{1})$ and $(\mathit{2})$ away from criticality, i.e., for $L \gg \xi$, and in absence of external field $H$ the free energy density $\cal F$ per volume $LL_\parallel^2$ and per $k_BT$ decomposes as (compare eq.~(\ref{free_away}))
\begin{eqnarray}
\fl {\cal F}^{(\mathit{1})}(t,H=0,L,L_\parallel) &= f_{\rm bulk}(t) + \frac{1}{L} f_{\rm surf}(t) + O(e^{-L/\xi}/L),\nonumber\\
\fl {\cal F}^{(\mathit{2})}(t,H=0,L,L_\parallel) &= f_{\rm bulk}(t) + \frac{1}{L} f_{\rm surf}(t) + \frac{\rho}{L^2} f_{\rm line}(t) + O(e^{-L/\xi}/L),
\label{extraline}
\end{eqnarray}
where $\rho=L/L_\parallel$ and $f_{\rm bulk}$, $f_{\rm surf}$, and $f_{\rm line}$ being independent of $L$ and $\rho$. In the limit $\rho\rightarrow 0$ the indicated correction terms in ${\cal F}^{(\mathit{1})}$ and ${\cal F}^{(\mathit{2})}$ reduce to the transversal finite size contributions $O(e^{-L/\xi}/L)$ which give rise to the critical Casimir forces in the lateral thermodynamic limit $L_\parallel\rightarrow\infty$. For large but finite values of $L_\parallel$ these correction terms acquire a dependence on $\rho$ which depends on the b.c. of the system.

Generalizing the arguments given in Ref.~\cite{Mon-89} and the corresponding discussion in \sref{sec:fss:energies}, we define the following quantity:
\begin{equation}
\label{edge}
\fl \hat{f}_{\rm line}(t,L)\equiv L^2\left(\frac{\partial}{\partial\rho}\Big|_{L,t}\left[{\cal F}^{(\mathit{2})}(t,H=0,L,L_\parallel)-{\cal F}^{(\mathit{1})}(t,H=0,L,L_\parallel)\right]\right)_{\big| \rho=0},
\end{equation}
so that, we have (compare with eq.~(\ref{surf_fp_sp_limit}))
\begin{equation}
\hat{f}_{\rm line}(t,L) = f_{\rm line}(t) + O(Le^{-L/\xi}),\qquad L\rightarrow\infty.
\end{equation}
Equation (\ref{edge}) formally defines a line free energy density $\hat{f}_{\rm line}$ also in the critical regime. If the slab $(\mathit{1})$ is realized as in the example given above, the free energy of this slab contains only bulk, surface, and finite size corrections which have at most a quadratic dependence on the aspect ratio $\rho$; in such a case the definition in eq.~(\ref{edge}) yields $\partial{\cal F}^{(\mathit{1})}/\partial\rho|_{\rho=0}=0$ and $\hat{f}_{\rm line}$ can be identified as the only free energy contribution of the two systems which varies $\propto \rho$ for small aspect ratios (see also the discussion below).

In the present case, we compare the free energy density for the geometry of a pair of individual chemical steps with the mean value of the free energy densities of the systems with laterally homogeneous b.c. $++$ and $+-$. For these latter b.c. in the critical region the free energy density decomposes into (compare eq.~(\ref{free_full_fss}) with $f=f_{++}$ or $f=f_{+-}$)
\begin{eqnarray}
\label{plusminus}
{\cal F}^{++}(t,H=0,L,\rho)&=f_{\rm bulk}^{\rm (ns)}(t) + \frac{1}{L}f_{\rm surf}^{\rm (ns)}(t) + \frac{1}{L^3}f_{++}(\tau,0,\rho),\nonumber\\
{\cal F}^{+-}(t,H=0,L,\rho)&=f_{\rm bulk}^{\rm (ns)}(t) + \frac{1}{L}f_{\rm surf}^{\rm (ns)}(t) + \frac{1}{L^3}f_{+-}(\tau,0,\rho),\nonumber\\
&\tau \equiv u_tL^{1/\nu},
\end{eqnarray}
where $f_{++}(\tau,0,\rho)$ and $f_{+-}(\tau,0,\rho)$ are scaling functions which describe the singular part of the free energy density for the laterally homogeneous $++$ and $+-$ b.c., respectively, in the absence of a bulk field\footnote{We note that in the absence of a bulk field, the surface free energy densities are the same for both b.c..}.
As in eq.~(\ref{edge}), the free energy density $\hat{f}_{\rm steps}$ for the pair of chemical steps is defined by
\begin{eqnarray}
\fl \hat{f}_{\rm steps}(t,L)\equiv L^2\Bigg(\frac{\partial}{\partial\rho}\Big|_{L,t}\Bigg[{\cal F}(t,H=0,L,\rho)\nonumber\\
-\frac{1}{2}{\cal F}^{++}(t,H=0,L,\rho)-\frac{1}{2}{\cal F}^{+-}(t,H=0,L,\rho)\Bigg]\Bigg)_{\big| \rho=0},
\label{csdef}
\end{eqnarray}
with ${\cal F}(t,L,\rho)$ as the free energy density for the system shown in \fref{lattice}. Its singular part $\hat{f}^{\rm (s)}_{\rm steps}$ is given by (see eq.~(\ref{plusminus}))
\begin{equation}
\label{cs}
\fl \hat{f}^{\rm (s)}_{\rm steps}(t,L) = \frac{1}{L}\left(\frac{\partial}{\partial\rho}\Big|_\tau\left[f(\tau,0,\rho)-\frac{1}{2}f_{++}(\tau,0,\rho)-\frac{1}{2}f_{+-}(\tau,0,\rho)\right]\right)_{\big| \rho=0},
\end{equation}
where $f(\tau,0,\rho)$ is the scaling function of the singular part of the free energy density for the system in the pair of individual chemical steps geometry and in absence of bulk field (see eq.~(\ref{free_full_fss})). In eqs.~(\ref{csdef}) and (\ref{cs}) we indicate with $++$ and $+-$ the quantities relative to the systems with laterally homogeneous $++$ and $+-$ b.c., while conforming to the notation of the previous sections the corresponding quantities for the system shown in \fref{lattice} are indicated without further specifications. Equation (\ref{cs}) renders the following relation, in lowest order in $\rho$, between the three functions $f(\tau,0,\rho)$, $f_{++}(\tau,0,\rho)$, and $f_{+-}(\tau,0,\rho)$:
\begin{eqnarray}
\fl f(\tau,0,\rho)=\frac{1}{2}f_{++}(\tau,0,\rho)+\frac{1}{2}f_{+-}(\tau,0,\rho) + \rho L \hat{f}^{\rm (s)}_{\rm steps}(t,L) + O(\rho^2),\quad \rho\rightarrow 0, T, L\ {\rm fixed}.\nonumber\\
\label{csrho}
\end{eqnarray}
Note that $L\hat{f}^{\rm (s)}_{\rm steps}(t,L) \ \vector(1,0){25}\hspace{-2em}^{L\gg\xi}\ \ Lf^{\rm (s)}_{\rm steps}(t)\sim L |t|^{2-\alpha_l}\sim |\tau|^\nu$, due to $\alpha_l=\alpha+2\nu$ \cite{Privman-89}, so that the r.h.s. of eq.~(\ref{csrho}) is indeed a function of $\rho$ and $\tau$. Together with the definition of the critical Casimir force in eq.~(\ref{casimir_def}) one has
\begin{equation}
\label{forces_decomposition}
F_C = \frac{1}{2}\left(F_{C,++} + F_{C,+-}\right) - \frac{\partial\left(\rho \hat{f}^{\rm (s)}_{\rm steps}(t,L)/L\right)}{\partial L}\Bigg|_{t,L_\parallel} + O(\rho^2),
\end{equation}
where
\begin{eqnarray}
&F_C=-\frac{\partial\left((1/L^2)f(\tau,0,\rho)-Lf^{\rm (s)}_{\rm bulk}(t)\right)}{\partial L}\Bigg|_{t,L_\parallel},\\
&F_{C,++}=-\frac{\partial\left((1/L^2)f_{++}(\tau,0,\rho)-Lf^{\rm (s)}_{\rm bulk}(t)\right)}{\partial L}\Bigg|_{t,L_\parallel},\\
&F_{C,+-}=-\frac{\partial\left((1/L^2)f_{+-}(\tau,0,\rho)-Lf^{\rm (s)}_{\rm bulk}(t)\right)}{\partial L}\Bigg|_{t,L_\parallel},
\end{eqnarray}
are the Casimir forces in the pair of individual chemical steps, $++$, and $+-$ geometry, respectively. The last term in eq.~(\ref{forces_decomposition}) is the contribution $F_{C,{\rm steps}}$ to the critical Casimir force due to the pair of individual chemical steps:
\begin{equation}
\label{forcecs}
\fl F_{C,{\rm steps}}=-\frac{\partial \left((\rho/L)\hat{f}^{\rm (s)}_{\rm steps}(t,L)\right)}{\partial L}\Bigg|_{t,L_\parallel} + O(\rho^2) = -\frac{1}{L_\parallel}\left(\frac{\partial \hat{f}^{\rm (s)}_{\rm steps}(t,L)}{\partial L}\Bigg|_{t,L_\parallel}\right) + O(\rho^2).
\end{equation}
According to the Monte Carlo results in Ref.~\cite{VGMD-08}, the dependence on $\rho$ of the critical Casimir force for the laterally homogeneous $++$ b.c. is negligible for $\rho\le 1/6$, while for the $+-$ b.c. it is quadratic and becomes relevant in the low-temperature phase. (Note that this quadratic term has a zero at $\tau=0$, see eq.~(\ref{taylor}).) This implies that eq.~(\ref{cs}) reduces to
\begin{equation}
\hat{f}^{\rm (s)}_{\rm steps}(t,L)=\frac{1}{L}\frac{\partial f(\tau,0,\rho)}{\partial\rho}\Bigg|_{\rho=0}.
\end{equation}
Inserting this result into eq.~(\ref{forcecs}) we obtain
\begin{eqnarray}
F_{C,{\rm steps}} &=-\frac{1}{L_\parallel}\left[\frac{\partial}{\partial L}\Bigg|_{t,L_\parallel}\left(\frac{1}{L}\frac{\partial f(\tau,0,\rho)}{\partial\rho}\Bigg|_{\rho=0}\right)\right] + O(\rho^2)\nonumber\\
&= \frac{\rho}{L^3}\left[\frac{\partial f(\tau,0,\rho)}{\partial\rho}\Bigg|_{\rho=0} -\frac{\tau}{\nu}\frac{\partial}{\partial \tau}\left(\frac{\partial f(\tau,0,\rho)}{\partial\rho}\Bigg|_{\rho=0} \right)\right]+O(\rho^2).
\label{intermediate}
\end{eqnarray}
On the other hand from $F_C=-\frac{\partial\left(Lf^{\rm (s)}_{\rm ex}\right)}{\partial L}|_{t,L_\parallel}$ (eq.~(\ref{casimir_def})) with $f^{\rm (s)}_{\rm ex}=\frac{1}{L^3}f-f^{\rm (s)}_{\rm bulk}$ (eq.~(\ref{free_ex_def})) one finds from eq.~(\ref{casimir_fss_leading})
\begin{equation}
\theta(\tau,\rho)=2f(\tau,0,\rho)+f^{\rm (s)}_{\rm bulk}(t)L^3-\frac{\tau}{\nu}\frac{\partial f(\tau,0,\rho)}{\partial \tau}-\rho\frac{\partial f(\tau,0,\rho)}{\partial \rho}
\end{equation}
so that
\begin{equation}
\frac{\partial\theta(\tau,\rho)}{\partial\rho}\Bigg|_{\rho=0}=\frac{\partial f(\tau,0,\rho)}{\partial\rho}\Bigg|_{\rho=0} -\frac{\tau}{\nu}\frac{\partial}{\partial \tau}\frac{\partial f(\tau,0,\rho)}{\partial\rho}\Bigg|_{\rho=0}.
\end{equation}
The comparison with eq.~(\ref{intermediate}) yields
\begin{equation}
F_{C,{\rm steps}} = \frac{\rho}{L^3}\frac{\partial\theta(\tau,\rho)}{\partial\rho}\Bigg|_{\rho=0} + O(\rho^2).
\end{equation}
With the expansion in eq.~(\ref{theta_rho}) we obtain
\begin{equation}
\label{result_Fstep}
F_{C,{\rm steps}}=\frac{\rho}{L^3}E(\tau) + O(\rho^2),\qquad \tau=t(L/\xi_0^+)^{1/\nu}.
\end{equation}
Since, as stated above, the dependence of $(F_{C,++}+F_{C,+-})/2$ on $\rho$ is quadratic, eq.~(\ref{result_Fstep}) implies that in the limit $\rho\rightarrow 0$ the contribution to $F_C$, which is linear in $\rho$, is solely due to the presence of the pair of individual chemical steps on one of the confining surfaces and thus serves as its fingerprint on the critical Casimir force.

We can extract this contribution from the MC data. From eqs.~(\ref{I_casimir_scaling}), (\ref{mod_rho}), and (\ref{theta_rho}) we have
\begin{equation}
\label{extract_D}
\frac{\partial I(\beta,L,L_\parallel)}{\partial\tilde{\rho}}\Bigg|_{\beta,L}=\frac{1}{(L-1/2+c)^3}E\left(t\left(\frac{L-1/2+c}{\xi_0^+}\right)^{1/\nu}\right).
\end{equation}
We note that the coupling parameter approach outlined in \sref{sec:simulations} results in the free energy difference between two systems with the same chemical steps. Therefore in $I(\beta,L,L_\parallel)$ non-singular background terms in the surface and line contributions to the free energy drop out. Since for every pair $\beta,L$ we have simulated systems of various aspect ratios, the derivative in eq.~(\ref{extract_D}) can be inferred from a simple fit linear in $\tilde{\rho}$, which is the same fit as the one which has been used in order to process the quantity $\tilde{I}(\beta,L,L_\parallel)$ in eq.~(\ref{I_tilde}). From this the function $E(\tau)$ follows according to eq.~(\ref{extract_D}).

\begin{figure}
\vspace{3em}
\begin{center}
\includegraphics[width=13cm,keepaspectratio]{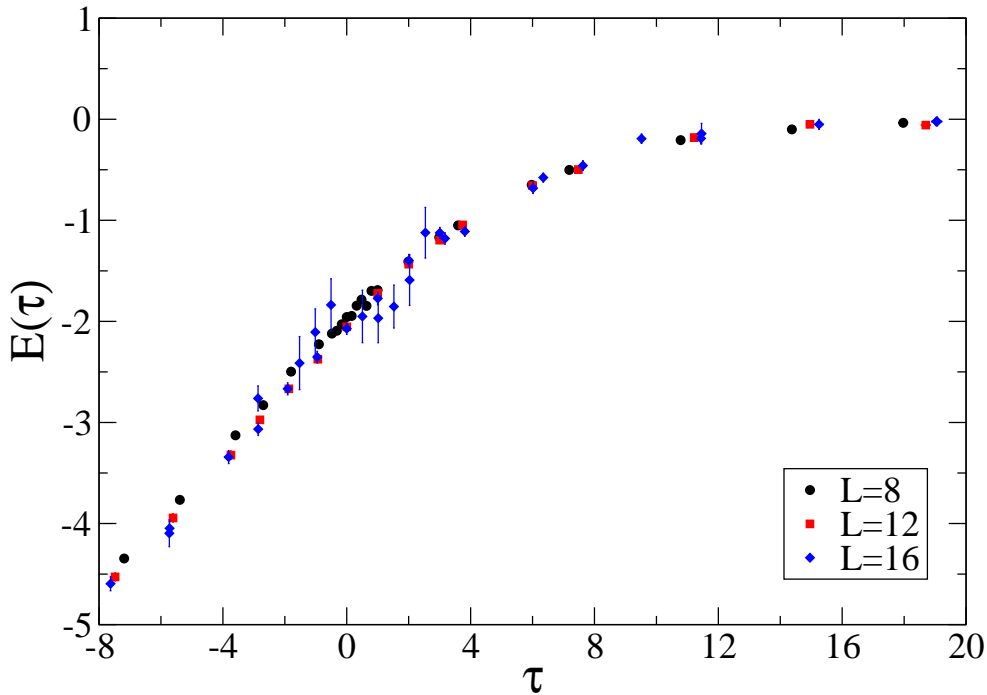}
\end{center}
\caption{The scaling function $E(\tau=t(L/\xi_0^+)^{1/\nu})$ which describes the contribution of the pair of individual chemical steps to the critical Casimir force via the dependence of the latter on the aspect ratio $\rho=L/L_\parallel$ (see eqs.~(\ref{theta_rho}), (\ref{forces_decomposition}), and (\ref{result_Fstep})). The statistical error bars for $L=8$ and $L=12$ are smaller than the symbol size.}
\label{der_rho}
\end{figure}

In \fref{der_rho} we show the function $E(\tau)$ as extracted from our MC data. It is negative within the whole range of $\tau$ values explored, which is consistent with the fact that the critical Casimir force is a decreasing function upon increasing $\rho$. The data for the various lattice sizes displayed in \fref{der_rho} collapse onto a single curve, with slight deviations for the data from the smallest lattice size. This is consistent with the
findings of \sref{sec:critical} and \sref{sec:theta:general}, according to which the data for $L=8$ are influenced by subleading scaling corrections. Moreover the data collapse confirms that the quantity we have extracted as the contribution of the pair of individual chemical steps indeed represents the singular part only, described by the scaling function $E(\tau)$ in eq.~(\ref{result_Fstep}). Since the critical Casimir force decays to $0$ if $|\tau|\rightarrow\infty$, which holds for an arbitrary aspect ratio $\rho$, it follows from eq.~(\ref{theta_rho}) that $E(\tau\rightarrow\pm\infty)=0$. This is confirmed in \fref{der_rho} for $\tau\rightarrow +\infty$, while the limiting behaviour for $\tau\rightarrow -\infty$ lies outside the available MC data. Therefore we expect that $E(\tau)$ reaches a minimum at $\tau=\tau_0<-8$.

According to the reconstruction scheme (see eq.~(\ref{gn})), in order to extract the scaling function $\theta(\tau,\rho)$ at some value $\tau\neq 0$ the system has to be simulated at $\left[B(\alpha,L)\right]^n\tau$, with $n$ given by the truncation of the sum in eq.~(\ref{gn}). In particular, for the present data one has $n=2$ and $B(\alpha,L)\simeq 3$ (see \sref{sec:theta:general}); thus in order to calculate $\theta(\tau,\rho)$ within the range $-8 \lesssim \tau \le 0$, one needs data for $-72 \lesssim \tau \le 0$, where the data in the interval $-72 \lesssim \tau \lesssim -8$ are used only to reconstruct the function $\theta(\tau,\rho)$. Although at sufficiently low temperatures the function $\theta(\tau,\rho)$ is suppressed, one still needs simulations for a large interval in the low-temperature phase. Since the computational cost of the simulations in the low-temperature phase increases with decreasing temperature, these circumstances limit the availability of data for $T<T_c$.

The film geometry studied here is relevant for the critical Casimir force in the presence of a chemically structured substrate. The simplest realization of such a substrate consists of a substrate which is finite in one lateral direction ($x$) and has a macroscopic extent in the other direction ($y$) so that it is de facto translationally invariant in this latter direction, exhibiting a single chemical step in $x$ direction (see \fref{lattice}). For such a substrate the film geometry can be approximately realized by either considering a wetting film of a binary liquid mixture \cite{he4,cwetting} or by a colloidal particle in front of such substrate \cite{SZHHB-08}. In the first case the wetting film thicknesses forming next to a $+$ or $-$ surface adjust to the different corresponding substrate potentials, so that the resulting critical Casimir force is described by the film geometry considered here only if the concomitant non-uniformity of the film thickness is small. This is conceivable because the wetting film thickness is mainly determined by the total density whereas the critical Casimir force is linked to the concentration fluctuations. In the second case the film geometry is approximately recovered if the radius of the colloidal particle is much larger than its distance from the substrate. In both cases the critical Casimir force is influenced by the lateral b.c. which in a first approximation could be treated as open ones. As we already mentioned in \sref{sec:intro}, for such a geometry the aspect-ratio dependence of the critical Casimir force is due to the presence of the chemical step and of the lateral edges, the contributions of which in general cannot be disentagled. On the other hand, the linear aspect-ratio dependence of the critical Casimir force calculated in this section is due to the presence of two individual chemical steps, while the corresponding contribution due to lateral edges could be determined in a similar way by considering a film geometry in the presence of homogeneous surfaces and laterally open b.c., or b.c. adapted to the actual experimental conditions realized there. Then the expected critical Casimir force $F_C$ for the film geometry in the presence of a single chemical step and laterally open b.c. (or b.c. adapted to the actual experimental conditions) is given by\pagebreak
\begin{eqnarray}
F_C = \frac{1}{L^3}\left[\theta(\tau,0) + \rho E_{\rm edges}(\tau)+ \frac{\rho}{2}E(\tau) + \rho \delta E(\tau)\right] + O(\rho^2), \nonumber\\
\quad \tau \equiv \left(\frac{T-T_c}{T_c}\right)\left(\frac{L}{\xi_0}\right)^{\frac{1}{\nu}},
\end{eqnarray}
where $E_{\rm edges}(\tau)$ is the scaling function associated with the aspect-ratio dependence due to the lateral edges, to be determined as described above, $\theta(\tau,\rho)$ and $E(\tau)$ are the scaling funtions calculated here, and $\delta E(\tau)$ is a scaling function which accounts for the expected non-additivity of the aspect-ratio dependence of the critical Casimir force with respect to the individual line contributions. One might expect that $\delta E(\tau)$ is small compared with the other three contributions.

\section{Order parameter profiles}
\label{sec:profiles}
The free energy and the critical Casimir forces provide integral informations about finite sized systems. Order parameter distributions deliver valuable additional and spatially resolved informations, which provide a deeper understanding and predictions which can be probed experimentally, e.g., by X-ray scattering under grazing incidence (see, e.g., Ref.~\cite{DH-95}).

We have computed the order parameter profiles for the system shown in \fref{lattice} at the critical temperature, for lattice sizes $L=16$, $24$, $32$, $40$, $48$ and aspect ratios $\rho=1/6$, $1/8$, $1/10$, $1/12$. We have sampled the order parameter in a region close to the ``central'' chemical step located at $z=x=0$ (see \sref{sec:simulations}), so that the presence of the ``lateral'' chemical step induced by the lateral periodicity is not relevant for the results here. In the following by referring to a chemical step we mean the central one. Certain details of the simulations are reported in \ref{sec:mc}. The leading scaling behaviour of the order parameter profiles $\Phi(x,z,L,L_\parallel)$ at criticality can be obtained from eq.~(\ref{scaling_mxz2}) by setting $t=0$:
\begin{equation}
\label{leading_m_ansatz}
\Phi(x,z,L,L_\parallel) = B \left(\frac{L}{\xi_0^\pm}\right)^{-\beta/\nu} \phi_{\rm c}\left(\frac{x}{L}, \frac{z}{L}, \rho\right),
\end{equation}
with (see eq.~(\ref{scaling_mxz3}))
\begin{equation}
\phi_{\rm c}\left(\hat{x}, \hat{z}, \rho\right)\equiv \phi_\pm\left(\hat{x}, \hat{z}, 0,\rho\right),
\end{equation}
where we have introduced the scaling function $\phi_{\rm c}\left(\hat{x},\hat{z},\rho\right)$ in order to simplify the notation. In addition to the leading scaling behaviour as of eq.~(\ref{leading_m_ansatz}), we observe scaling corrections $\propto 1/L$. However, within the precision of our data, no aspect ratio dependence has been found in the spatial region near the chemical step which we have considered here. In order to extract the thermodynamic limit, we fit our MC data for $\Phi(x,z,L,L_\parallel)$ to the following expression:
\begin{eqnarray}
\label{m_ansatz}
\Phi(x,z,L,L_\parallel) = B \left(\frac{L}{\xi_0^\pm}\right)^{-\beta/\nu}\left(\phi_{\rm c}\left(\frac{x}{L}, \frac{z}{L}\right) + \frac{1}{L}g\left(\frac{x}{L}, \frac{z}{L}\right)\right),
\end{eqnarray}
where, with a slight abuse of notation, we have dropped the dependence on the aspect ratio $\rho$, and $\phi_{\rm c}$ and $g$ are unconstrained functions. This is achieved by performing a fit linear in $1/L$ of $\Phi(x,z,L)L^{\beta/\nu}$ for every pair $(x/L,z/L)$, using $\beta/\nu=(1+\eta)/2=0.51819(8)$\cite{CPRV-02}. Error bars have been determined from the jackknife procedure (see, e.g., Ref.~\cite{AM-book}) in order to take into account the statistical covariance of the data sampled from the same MC run for various pairs $(x/L,z/L)$. The non-universal amplitude combination appearing in eq.~(\ref{m_ansatz}) is given by (see eqs.~(\ref{scaling_mxz}) and (\ref{spontm}))
\begin{equation}
\label{m_normalization}
B(\xi_0^+)^{\beta/\nu} = 0.933(6)
\end{equation}
and has been computed by analyzing the $25^{\rm th}$-order high-temperature expansion reported in Ref.~\cite{CPRV-02} (see \ref{sec:ht} for details).
In sections \ref{sec:critical} and \ref{sec:theta} we observed  that data for the lattice size $L=8$ are affected by next-to-leading scaling corrections, while for $L\ge 10$ no subleading scaling corrections are observed. In the present data for the order parameter profiles the minimum lattice size is $L=16$, so that corrections to eq.~(\ref{m_ansatz}) should be negligible. In fact, our fits always render a good $\chi^2/DOF$, except for a narrow spatial region close to the chemical step. This indicates that there effects due to subleading corrections are probably strong. In fact, if scaling corrections are (at least partially) due to analytic corrections of the scaling variables as discussed in \sref{sec:fss}, the function $g(x/L,z/L)$ should be expressable in terms of derivatives of the function $\phi_{\rm c}(x/L,z/L)$ (compare with eq.~(\ref{corr_scaling_casimir})). Since close to the chemical step the function $\phi_{\rm c}(x/L,z/L)$ varies steeply, the ensuing leading correction to scaling $\propto L^{-1}$ can potentially become large enough as to invalidate the ansatz given in eq.~(\ref{m_normalization}).

\begin{figure}
\vspace{3em}
\begin{center}
\includegraphics[width=30em,keepaspectratio]{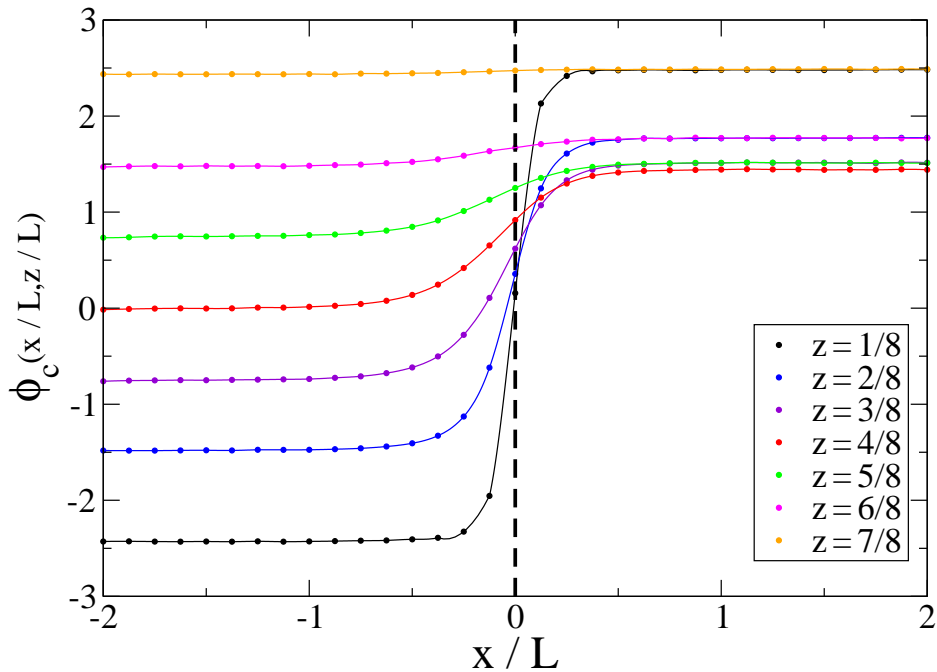}
\end{center}
\caption{The universal order parameter scaling function $\phi_{\rm c}(x/L,z/L)$ at criticality as defined by eq.~(\protect\ref{leading_m_ansatz}). For $x<0$ ($x\ge 0$) there are $-$ ($+$) b.c. on the lower confining surface at $z=0$ whereas there are $+$ b.c. on the upper surface at $z=L-1$.
The error bars are smaller than the symbol size. The data points are interpolated with a cubic spline.}
\label{m_profiles}
\end{figure}

In \fref{m_profiles} we report the function $\phi_{\rm c}\left(x/L, z/L\right)$, for various values of $z/L$ in the region close to the chemical step located at $x=0$. For our simulations we have chosen the coordinate system such that the confining surface with the spins fixed to form a chemical step corresponds to $z=0$, while the homogeneous surface with the spins fixed to $+1$ is located at $z=L-1$. In the lateral direction the origin $x=0$ corresponds, in the lower surface, to spins fixed to $+1$, whose left neighbours are spins fixed to $-1$\footnote{In the finite size scaling limit, i.e., in the limit $L\rightarrow\infty$ at fixed $z/L,x/L$, the precise choice of the coordinate system (e.g., whether the lower boundary is at $z=0$ or $z=1$) does not affect $\phi_{\rm c}(x/L,z/L)$; it only influences the correction-to-scaling function $g\left(x/L, z/L\right)$.}.

By inspecting the profiles we see that close to the chemical step at $|x/L|\ll 1$ and $z/L=1/8$, the function varies strongly, passing from negative values at $x/L<0$ to positive values  at $x/L>0$. As expected, upon increasing $z/L$ the gradient at $x/L=0$ decreases, as the frustration due to the chemical step heals. At $z/L=7/8$, i.e., close to the laterally homogeneous boundary, the scaling function is almost flat, signalling that there the effect due to the presence of the chemical step on the distant wall is small. We also note that for $|x/L| \gtrsim 0.5$, there is de facto no dependence of the order parameter profiles on the lateral coordinate $x$: at such distances from the chemical step, the order parameter is no longer influenced by the chemical step and we recover the profiles (as a function of the coordinate $z$ normal to the surface) for homogeneous boundaries: $-+$ for $x/L\lesssim -0.5$ and $++$ for $x/L\gtrsim 0.5$. Consistently, for $x/L\gtrsim 0.5$ the scaling function is antisymmetric around $z/L=1/2$ and for $x/L\gtrsim 0.5$ it is symmetric around $z/L=1/2$. These findings support the line of reasoning at the end of \sref{sec:simulations}, concerning the various contributions to the critical Casimir force, even at $T_c$.

Close to the walls and in the regions where the profiles resemble those for laterally homogeneous b.c., the scaling of the order parameter is predicted to exhibit the distant-wall corrections described by eqs.~(\ref{decay}), (\ref{wallcorr}), and (\ref{wallcorr2}). We first consider the case of $++$ b.c.. To this end we analyze the data close to the bottom wall and for $x/L\ge 1$ in order to avoid a potential bias due to a residual dependence on $x/L$. We fit them to the expression
\begin{equation}
\label{fit_profile}
\Phi(x,z,L)=CL^{-\beta/\nu}\left(\frac{z}{L}\right)^{-\beta/\nu}\left[1+A\left(\frac{z}{L}\right)^3\right]\left(1+\frac{1}{L}g(z/L)\right),
\end{equation}
with $C$, $A$, and $g(z/L)$ as free parameters. The correction term $\sim (z/L)^3$ is expected to be valid only if $z/L$ is sufficiently small. Thus we have performed the fits by considering only those data with $(z/L)\le(z/L)_{\rm max}$. The corresponding results are reported in \tref{res_fit_profile}.

\Table{\label{res_fit_profile}Fits of the order parameter profile in the region with $++$ b.c., i.e., $x/L\ge 1$ and $(z/L)\le(z/L)_{\rm max}$ to eq.~(\protect\ref{fit_profile}). The indicated error bars are the sum of the statistical uncertainty (first number, obtained with the jackknife technique \protect\cite{AM-book}) and of the variation of $\beta/\nu=(1+\eta)/2=0.51819(8)$\protect\cite{CPRV-02} due to its error bar (second number).}
\begin{tabular}{@{}lllll}
\br
$(z/L)_{\rm max}$ & $C$ & $A$ & $g(z/L)$ & $\chi^2/DOF$\\
\mr
$3/8$              & $0.7876(2+1)$ & $1.442(7+2)$   & $g(1/8)=-1.825(3+2)$ & $\0 1.24$\\
                   &               &                & $g(2/8)=-0.798(4+1)$ &\\
                   &               &                & $g(3/8)=-0.338(8+2)$ &\\
\mr
$4/8$              & $0.7878(2+2)$ & $1.408(3+1)$   & $g(1/8)=-1.831(4+2)$ & $\0 1.26$\\
                   &               &                & $g(2/8)=-0.795(4+1)$ &\\
                   &               &                & $g(3/8)=-0.309(5+1)$ &\\
                   &               &                & $g(4/8)=\m 0.060(9+2)$ &\\
\mr
$5/8$              & $0.7881(2+2)$ & $1.393(2+1)$ & $g(1/8)=-1.837(4+3)$ & $\0 1.24$\\
                   &               &              & $g(2/8)=-0.796(4+1)$ &\\
                   &               &              & $g(3/8)=-0.300(4+1)$ &\\
                   &               &              & $g(4/8)=\m 0.089(5+2)$ &\\
                   &               &              & $g(5/8)=\m 0.456(8+3)$ &\\
\mr
$6/8$              & $0.7878(2+2)$ & $1.4015(7+3)$ & $g(1/8)=-1.831(4+2)$ & $\0 1.26$\\
                   &               &               & $g(2/8)=-0.792(4+2)$ &\\
                   &               &               & $g(3/8)=-0.302(4+1)$ &\\
                   &               &               & $g(4/8)=\m 0.076(4+1)$ &\\
                   &               &               & $g(5/8)=\m 0.423(4+2)$ &\\
                   &               &               & $g(6/8)=\m 1.019(6+2)$ &\\
\mr
$7/8$              & $0.7810(1)$ & $1.5146(3)$ & $g(1/8)=-1.662(4)$ & $56.7$\\
                   &             &             & $g(2/8)=-0.654(4)$ &\\
                   &             &             & $g(3/8)=-0.247(4)$ &\\
                   &             &             & $g(4/8)=-0.031(4)$ &\\
                   &             &             & $g(5/8)=\m 0.051(3)$ &\\
                   &             &             & $g(6/8)=\m 0.267(2)$ &\\
                   &             &             & $g(7/8)=\m 2.268(2)$ &\\
\br
\end{tabular}
\endTable

By analyzing conservatively the dependence of the results of the fit on $(z/L)_{\rm max}$, we obtain the following estimates:
\begin{eqnarray}
\label{estimate_C}
C=0.7879(5),\\
\label{estimate_A}
A=1.40(1),\\
g(1/8)=-1.834(8),\\
g(2/8)=-0.794(6),\\
g(3/8)=-0.305(9),\\
g(4/8)=0.07(2),\\
g(5/8)=0.44(2).
\end{eqnarray}
By comparing eq.~(\ref{fit_profile}) with eqs.~(\ref{wallcorr}), (\ref{wallcorr2}), and (\ref{decay}) one has $C=B(\xi_0^+)^{\beta/\nu} c_+$. By inserting the estimates given in eqs.~(\ref{m_normalization}) and ~(\ref{estimate_C}) we obtain
\begin{equation}
\label{estimate_cplus}
c_+=0.844(6).
\end{equation}
We can compare this result with previous theoretical estimates $c_+=0.87(7)$ \cite{SDL-94}, $c_+=0.94(5)$ \cite{FD-95}, and $c_+=0.857$\cite{BU-01}. From a variety of experimental data corresponding to different liquids one infers $c_+=1.60(42)$, $c_+=0.77(19)$, $c_+=1.14(29)$, $c_+=0.91(26)$, $c_+=1.05(9)$, $c_+=1.02(10)$, $c_+=1.25(9)$, $c_+=0.84(15)$ \cite{FD-95}. Our result is one order of magnitude more precise than previous theoretical determinations and in very good agreement with the results in Refs.~\cite{SDL-94,BU-01}, while the discrepancy with Ref.~\cite{FD-95} is of only two error bars. The comparison with the experimental data extracted in Ref.~\cite{FD-95} is less satisfactory: half of the experimental data are in agreement with our result, while the apparently more precise determinations are not compatible with our result.

The comparison of eq.~(\ref{fit_profile}) with eqs.~(\ref{wallcorr}), (\ref{wallcorr2}), and (\ref{decay}) yields \mbox{$A=-(d-1)\Delta_{++}C_+$}, where $\Delta_{++}$ is the critical Casimir amplitude for $++$ b.c., and $C_+$ is a universal coefficient which depends only on the surface universality class of the close surface, i.e., here the extraordinary or normal surface universality class. With $(d-1)\Delta_{++}=-0.820(15)$ \cite{Hasenbusch-10c} we obtain
\begin{equation}
\label{estimate_Cex}
C_+=1.71(4).
\end{equation}

This coefficient $C_+$ has been determined for the extraordinary surface universality class in Ref.~\cite{ES-94}, using a combination of the $\varepsilon$-expansion technique and interpolation with the exact $d=2$ result leading to $C_+ \sim 1.42 - 1.96$. Our determination is more precise, and in full agreement with this estimate.

We have repeated the above procedure for the data of the profile for $x/L\le -1$, corresponding to $+-$ b.c. In this case the fits of eq.~(\ref{fit_profile}) result in a large $\chi^2/DOF$. Due to $(d-1)\Delta_{+-}=5.613(20)$ \cite{Hasenbusch-10c}, in this case the distant wall gives rise to a correction which is about seven times stronger that in the $++$ case. Using the estimate in eq.~(\ref{estimate_Cex}) and the Casimir amplitudes $(d-1)\Delta_{++}=-0.820(15)$ and $(d-1)\Delta_{+-}=5.613(20)$ \cite{Hasenbusch-10c}, we can infer that eq.~(\ref{wallcorr2}) is presumably valid for distances $z$ such that
\begin{eqnarray}
&C_+(d-1)\Delta_{+\pm}\left(\frac{z}{L}\right)^3 \ll 1,\\
\label{bound_pp}
&\frac{z}{L} \ll 0.9,\qquad {\rm for} ++\ {\rm b.c.},\\
\label{bound_pm}
&\frac{z}{L} \ll 0.5,\qquad {\rm for} +-\ {\rm b.c.}.
\end{eqnarray}
Equation (\ref{bound_pm}) explains why fits to eq.~(\ref{fit_profile}) fail for $+-$ b.c. for the presently available MC data. In fact, only the data for $z/L=1/8$ and $2/8$ can satisfy eq.~(\ref{bound_pm}), but they are insufficient for performing a fit to eq.~(\ref{fit_profile}). On the other hand, the bound given by eq.~(\ref{bound_pp}) is consistent with the results in \tref{res_fit_profile}, where fits to eq.~(\ref{fit_profile}) turn out to be reliable up to $(z/L)_{\rm max}=6/8=0.75$.

\section{Summary and conclusions}
\label{sec:conclusions}
We have studied the critical Casimir force and the order parameter profiles for a three-dimensional slab of thickness $L$ belonging to the Ising universality class. We have employed periodic boundary conditions in the two lateral directions of extents $L_\parallel$ and fixed boundary conditions on the two confining surfaces. The Ising spins on the upper surface are fixed to $+1$. The lower surface is divided into two halves, one with spins fixed to $-1$ and the other with spins fixed to $+1$ (see \fref{lattice}). We have investigated this system by combining Monte Carlo (MC) simulations and finite-size scaling analysis.
We have obtained the following main results:
\begin{itemize}
\item In the finite-size scaling limit $t=(T-T_c)/T_c\rightarrow 0$, $L\rightarrow\infty$, with $\xi/L$ and $L/L_\parallel$ fixed, the singular part of the critical Casimir force $F_C$ per area $L_\parallel^2$ and per $k_BT$ takes on the universal scaling form $F_C(t,L,L_\parallel)=L^{-3}\theta(\tau=t(L/\xi_0^+)^{1/\nu},\rho=L/L_\parallel)$ (eq.~(\ref{casimir_fss_leading})) with $\xi(t\rightarrow 0^\pm)=\xi_0^\pm|t|^{-\nu}$ as the true bulk correlation length and $\rho$ as the aspect ratio of the slab. The force $F_C$ is determined by integrating the thermal Monte Carlo average of a suitable crossover Hamiltonian (see equations (\ref{I_def}), (\ref{I_casimir}), and (\ref{I_casimir_scaling})).
\item In order to be able to extract the universal scaling function $\theta(\tau, \rho)$ from the MC data for a finite size system, particular care has been taken to minimize the influence of corrections to scaling. To this end we have studied a suitable Blume-Capel model which belongs to the Ising universality class and which suppresses the leading correction to scaling $\propto L^{-0.832}$ (eq.~(\ref{BC})). From the available high-temperature series of this model we have determined the relevant non-universal amplitudes (\ref{sec:ht}). This allows one to properly normalize the scaling functions and scaling variables. A detailed finite-size scaling analysis leads to a relation between the actual MC observable called $I$ (eq.~(\ref{I_def})) and the desired scaling function $\theta(\tau,\rho)$ with suitably adjusted scaling variables in order to minimize corrections to scaling (eq.~(\ref{I_casimir_scaling})). This procedure yields numerically accurate data for the scaling function $\theta(\tau,\rho)$.
\item As expected on general grounds (\fref{fig_decomposition}), in the limit of vanishing aspect ratio $\rho=L/L_\parallel\rightarrow 0$ the critical Casimir force for the system shown in \fref{lattice} reduces to the mean value of the critical Casimir forces for laterally homogeneous $++$ and $+-$ boundary conditions so that $\theta(\tau,\rho\rightarrow 0)=\frac{1}{2}[\theta_{++}(\tau)+\theta_{+-}(\tau)]$ (see \fref{thetar0} and $\rho\rightarrow 0$ and mv(IV) in \fref{thetacomparison}). Since $F_{C,+-}$ is more repulsive than $F_{C,++}$ is attractive, $\theta(\tau,\rho\rightarrow 0)$ is positive (figures~\ref{thetar0} and \ref{thetacomparison}) so that for $\rho\rightarrow 0$ the critical Casimir force for the system in \fref{lattice} is repulsive.
\item The presence of the pair of individual chemical steps documents itself in the dependence of the scaling function $\theta(\tau,\rho)$ on the aspect ratio $\rho$. For laterally homogeneous b.c. in the limit $\rho\rightarrow 0$ the critical Casimir force generally approaches its limiting value $\propto \rho^2$. The amplitude of this quadratic term vanishes at $T_c$ (see eq.~(\ref{taylor})). The presence of the pair of individual chemical steps generates an additional line contribution $\propto L_\parallel$ to the free energy (eq.~(\ref{extraline})) which causes a leading, linear dependence on $\rho$ (\fref{casimir_amplitude}). Based on a suitable extrapolation scheme, at $T_c$ this linear variation of $\theta(\tau=0,\rho)=\Theta(\rho)$ is shown in \fref{critvsrho}. Since for $\rho\rightarrow 0$ the linear dependence on $\rho$ dominates, the decrease of $\Theta(\rho)$ upon increasing $\rho$ implies that the presence of the chemical steps weakens the repulsive critical Casimir force relative to the mean value of the forces for the corresponding laterally homogeneous b.c..
\item For $T\neq T_c$ we have determined the scaling function $\theta(\tau,\rho)$ for $\rho=1/6$, $1/8$, $1/10$, $1/12$ (figure~\ref{thetar6}, \ref{thetar8}, \ref{thetar10}, and \ref{thetar12}, respectively). It turns out that $\theta(\tau,\rho)<\theta(\tau,\rho=0)$ for the available values of the scaling variables $\tau$ with the deviation being most pronounced for $\tau<0$ (\fref{thetacomparison}). This difference is captured by the scaling function $E(\tau)$ defined via $\theta(\tau,\rho\rightarrow 0) = \theta(\tau,\rho=0) + \rho E(\tau) + O(\rho^2)$ with $E(\tau)<0$ and $E(\tau\rightarrow\pm\infty)=0$ (see \fref{der_rho} and the discussion at the end of \sref{sec:theta:step}). This linear contribution $\sim E(\tau)$ is solely due to the pair of individual chemical steps. Also for this laterally inhomogeneous system the amplitude of the quadratic term vanishes at $T_c$ (see eq.~(\ref{taylor})).
\item At the critical temperature we have determined the order parameter profiles for the system shown in \fref{lattice} (see \fref{m_profiles}). Since in the spatial region around the chemical step which we have considered here the dependence of the order parameter profiles on the aspect ratio is weak, these profiles can be described by the scaling function $\phi_c(x/L,z/L,\rho=0)$ (see eq.~(\ref{leading_m_ansatz})). Sufficiently away from the chemical step the order parameter profiles reduce to the ones which correspond to the laterally homogeneous b.c. $++$ or $+-$. For the latter ones we have determined the universal amplitude $c_+$ (equation (\ref{estimate_cplus})) characterizing the leading behaviour of the order parameter near a wall (equation (\ref{decay})) for which experimental data are available \cite{FD-95}, as well as the amplitude of the leading distant wall correction (equations (\ref{wallcorr}) and (\ref{wallcorr2}), and (\ref{estimate_Cex})), which is in full agreement with previous, but less accurate, estimates.
\end{itemize}

Our results are relevant for the critical behaviour of confined systems belonging to the Ising universality class and in the presence of a chemically structured substrate. Indeed, such a system has been experimentally realized; the critical Casimir force has been probed by a spherical colloidal particle close to a structured substrate \cite{SZHHB-08,NHB-09}.
The film geometry studied here is realized approximately by such a system if the radius of the colloid is large compared to its distance from the substrate \cite{TKGHD-09}. Another possibility would be to monitor the thickness of a wetting film of a classical binary liquid mixture near its critical end point of demixing in equilibrium with its vapour phase \cite{cwetting} and in contact with a chemically structured substrate. However, in such a system in addition the wetting film thicknesses forming next to a $+$ or $-$ surface adjust to different adsorption preference. This adds a new interesting aspect to the problem. Some additional considerations regarding a possible experimental realization are reported at the end of \sref{sec:theta:step}.

The present study points towards several interesting issues to be investigated in the future. A natural generalization consists of studying the critical Casimir force in the presence of many stripes with alternating adsorption preferences. In this case the critical Casimir force depends additionally on the width of the stripes (see Ref.~\cite{TKGHD-09} where the critical Casimir force between such a substrate and a colloid has been studied; genuine three-dimensional simulation data for this system are still missing). Furthermore, one can consider an alternating adsorption preference also on the upper boundary, in which case even for a slab also a lateral Casimir force arises. Since such a system has been studied within mean field theory in Ref.~\cite{SSD-06}, a comparison with three-dimensional simulation data would be of particular interest. The critical Casimir forces for more complicated geometries, such as the square-patterned substrate experimentally realized in Ref.~\cite{SZHHB-08}, have not yet been studied theoretically.

From a more theoretical point of view, the nature and the origin of the scaling corrections for non-periodic boundary conditions calls for further analysis. In particular, so far the ansatz of eq.~(\ref{casimir_fss}) appears to describe correctly those additional corrections to scaling proportional to $1/L$ which emerge in the presence of non-periodic boundary conditions. This has been also checked numerically in Refs.~\cite{Hasenbusch-08,Hasenbusch-09,Hasenbusch-09b} for the $XY$ model with free surfaces and in Ref.~\cite{Hasenbusch-10c} for the Ising universality class with fixed surfaces. A deeper theoretical understanding of this fact in terms of renormalization group theory would be highly welcome.

\ack
We are grateful to Ettore Vicari, Andrea Pelissetto and Martin Hasenbusch for useful discussions. We thank Oleg Vasilyev for providing us the Monte Carlo data of Ref.~\cite{VGMD-08}. Correspondence with Volker Dohm is gratefully acknowledged.
\appendix

\section{Analysis of the high-temperature series}
\label{sec:ht}
In this appendix we analyze the $25^{\rm th}$-order high-temperature expansion for the improved Blume-Capel model reported in Ref.~\cite{CPRV-02}, with the aim of calculating non-universal amplitudes associated with various observables, which allows us to properly normalize scaling variables and universal scaling functions. We follow closely the notation in Refs.~\cite{CPRV-02,PV-02}. We consider the magnetization $M$ per volume, the two-point correlation function $G(x)$ and the corresponding moments $m_{2j}$, and the four-point susceptibility $\chi_4$ defined as
\begin{eqnarray}
&M\equiv\frac{1}{V}\<\sum_xS(x)\>,\\
&G(x)\equiv \<S(0)S(x)\>, \\
&m_{2j}\equiv \sum_x |x|^{2j}G(x),\\
&\chi_4\equiv \sum_{x_1,x_2,x_3}\<S(0)S(x_1)S(x_2)S(x_3)\>_c,
\end{eqnarray}
where the subscript $c$ in the definition of $\chi_4$ indicates the connected part of the thermal average of the product of spins.
The susceptibility $\chi$ and the second-moment correlation length $\xi_{\rm 2nd}$ follow from
\begin{equation}
\chi=m_0,
\end{equation}
and
\begin{equation}
\xi_{\rm 2nd}^2=\frac{m_2}{6\chi}.
\end{equation}
Close to the critical temperature these quantities exhibit the following singular behaviours:
\begin{eqnarray}
\label{chi_amplitude}
\chi&=C^+t^{-\gamma}, &t>0,\\
\label{xi_amplitude}
\xi_{\rm 2nd}&=\xi_{0, {\rm 2nd}}^+ t^{-\nu}, &t>0,\\
\label{chi4_amplitude}
\chi_4&=-C_4^+t^{-\gamma_4}, &t>0\\
\label{M_amplitude}
M&=B(-t)^{-\beta}, \quad &t<0,
\end{eqnarray}
where $t\equiv (T-T_c)/T_c\rightarrow 0$ is the reduced temperature. We note that the exponential (or true) correlation length $\xi$, which governs the exponential decay of $G(x)$, diverges for $t\rightarrow 0^+$ as
\begin{equation}
\label{xi_gap}
\xi=\xi_0^+ t^{-\nu}, \quad t>0,
\end{equation}
where the ratio $\xi_0^+/\xi_{0, {\rm 2nd}}^+=1.000200(3)$ \cite{CPRV-02} is universal.

We analyze the high-temperature expansion using quasi-diagonal first- and second-order integral approximants (IA1 and IA2, respectively) \cite{Guttmann-89}. IA1 are the solutions $f(x)$ of the first-order differential equation
\begin{equation}
\label{IA1}
P_1(x)f'(x)+P_0(x)f(x)+R(x)=0,
\end{equation}
where $P_1(x)$, $P_0(x)$, and $R(x)$ are polynomials of order $m_1$, $m_0$, and $k$, respectively. The overall normalization of the equation is fixed by setting $P_1(0)=1$ and the coefficients of $P_1(x)$, $P_0(x)$, and $R(x)$ are fixed by the requirement that the Taylor expansion of $f(x=\beta\rightarrow 0)$ matches the known high-temperature expansion. If $x_{c1}$ is the smallest real positive root of $P_1(x)$, for $x\rightarrow x_{c1}$ the solution of eq.~(\ref{IA1}) behaves as
\begin{eqnarray}
\label{IA1sol}
f(x)&\simeq A_1(x)|x-x_{c1}|^{e_1}+B_1(x),\qquad x\rightarrow x_{c1},\nonumber\\
&e_1=-\frac{P_0(x_{c1})}{P_1'(x_{c1})},\nonumber\\
B_1(x_{c1})&=-\frac{R(x_{c1})}{P_0(x_{c1})},
\end{eqnarray}
where, near $x=x_{c1}$, $A_1(x)$ and $B_1(x)$ are analytic functions which are determined by $P_0(x)$, $P_1(x)$, and $R(x)$.

IA2 are solutions of the second-order differential equation
\begin{equation}
\label{IA2}
P_2(x)f''(x)+P_1(x)f'(x)+P_0(x)f(x)+R(x)=0,
\end{equation}
where $P_2(x)$, $P_1(x)$, $P_0(x)$, and $R(x)$ are polynomials of order $m_2$, $m_1$, $m_0$, and $k$, respectively, and $P_2(0)=1$. Again, the coefficients of the polynomials in eq.~(\ref{IA2}) are determined by the high-temperature expansion of $f(x)$. The solution of eq.~(\ref{IA2}) has a structure which is similar to the one in eq.~(\ref{IA1sol}): close the smallest real positive root $x_{c2}$ of $P_2(x)$ one has
\begin{eqnarray}
\label{IA2sol}
f(x)&\simeq A_2(x)|x-x_{c2}|^{e_2}+B_2(x),\qquad x\rightarrow x_{c2},\nonumber\\
&e_2=-\frac{P_1(x_{c2})}{P_2'(x_{c2})},\nonumber\\
B_2(x_{c2})&=-\frac{R(x_{c2})}{P_0(x_{c2})}.
\end{eqnarray}
Inspection of eqs.~(\ref{IA1sol}) and (\ref{IA2sol}) tells that the critical exponents $e_1$ and $e_2$ and the background terms $B_1(x_{c1})$ and $B_2(x_{c2})$ can be obtained without solving explicitly eqs.~(\ref{IA1}) and (\ref{IA2}), because it is sufficient to determine the roots of $P_1(x)$ and $P_2(x)$, respectively.

However, we are interested in the non-universal amplitudes $A_i(x)$, which cannot be extracted directly from the coefficients of the polynomials $P_i(x)$ and $R(x)$. In order to obtain them, we use the method outlined in Ref.~\cite{LF-89}. Given the high-temperature series for a quantity $Q(\beta)$ which close to criticality diverges as $Q(\beta)\simeq A|\beta_c/\beta-1|^{-\alpha}$, we analyze the series of $\tilde{Q}(\beta)\equiv Q(\beta)(1-\beta/\beta_c)^\alpha$. The background term obtained from the analysis of $\tilde{Q}(\beta)$ corresponds to the desired amplitude $A(\beta)$ of the quantity $Q(\beta)$.
This approach requires the knowledge of the critical inverse temperature $\beta_c$ as well as of the critical exponent $\alpha$ which characterizes the critical behaviour of the given quantity $Q(\beta)$. As in Ref.~\cite{CPRV-02}, we consider {\it quasi-diagonal} approximants IA1 and IA2, i.e., integral approximants constructed from polynomials $P_i(x)$, $R(x)$ of almost equal degree, which are expected to lead to more reliable results \cite{Guttmann-89}. If the series expansion of $Q(\beta)$ is known up to the order $\beta^n$, we limit the orders of the polynomials $P_i(x)$ and $R(x)$ such that in the case of IA1 we have
\begin{equation}
\label{diagonalIA1}
{\rm Max}\{(n-2)/3-q,2\} \le m_0,m_1,k \le (n-2)/3+q,
\end{equation}
and in the case of IA2
\begin{equation}
\label{diagonalIA2}
{\rm Max}\{(n-4)/4-q,2\} \le m_0,m_1,m_2,k \le (n-4)/4+q.
\end{equation}
In eqs.~(\ref{diagonalIA1}) and (\ref{diagonalIA2}) $q$ measures the off-diagonality allowed, i.e., it limits the difference between the degrees of the polynomials $P_i(x)$ and $R(x)$. We have considered $q=3$ for IA1 and $q=2$ for IA2. In order to improve the reliability of the result, we have always considered \emph{biased} approximants, i.e., we set $P_1(x)=(1-x/\beta_c)\tilde{P}_1(x)$ in eq.~(\ref{IA1}) and $P_2(x)=(1-x/\beta_c)\tilde{P}_2(x)$ in eq.~(\ref{IA2}), so that the singularity occurs at $x=\beta_c$. Furthermore, as in Refs.~\cite{CPRV-99,CPRV-02} we discard the approximants which lead to spurious singularities in the region of the complex plane given by
\begin{equation}
\label{spurious}
x_{\rm min}\le {\rm Re}\ z \le x_{\rm max},\qquad |{\rm Im}\ z| \le y_{\rm max},
\end{equation}
where $z\equiv \beta/\beta_c$ and the size of the rectangle is chosen as in Ref.~\cite{CPRV-99} as $x_{\rm min}=0.5$, $x_{\rm max}=1.5$, and $y_{\rm max}=0.5$, where this choice was motivated by considering stability criteria.

In Ref.~\cite{CPRV-02} the high-temperature series of $\chi(\beta)$ for the improved Blume-Capel model is reported up to the $25^{\rm th}$ order. With the method described above, we have determined the non-universal amplitude $C^+$ appearing in eq.~(\ref{chi_amplitude}). With IA1 approximants we obtain
\begin{equation}
\label{chi_IA1}
C^+=0.465994(3)\pm  3\times 10^{-4} \pm 6\times 10^{-4},\qquad {\rm for~IA1,}
\end{equation}
where the first quoted error indicates the spread of approximants (standard deviation), the second stems from the variation of $\beta_c=0.3856717(10)$ \cite{CPRV-02} within one error bar, and the third one originates from the error bar of $\gamma=1.2373(2)$ \cite{CPRV-02}. With IA2 approximants we obtain
\begin{equation}
\label{chi_IA2}
C^+=0.4664(7) \pm 4 \times 10^{-4} \pm 6 \times 10^{-4},\qquad {\rm for~IA2.}
\end{equation}
As a final estimate we take $C^+=0.466(2)$.

The analysis of the $21^{\rm th}$-order series of $\chi_4$ reported in Ref.~\cite{CPRV-02} gives the following results:
\begin{equation}
C_4^+=0.3683(7)\pm 3\times 10^{-4} \pm 10^{-3}, \qquad {\rm for\ IA1},
\end{equation}
and
\begin{equation}
C_4^+=0.365(4)\pm 6\times 10^{-4}\pm 10^{-3}, \qquad {\rm for\ IA2},
\end{equation}
where again the first error indicates the spread of approximants, the second one is due to the uncertainty of $\beta_c$, and the third one follows from the spread of $\gamma_4=2\gamma+3\nu=4.3650(6)$ \cite{CPRV-02}. As a final estimate we take $C_4^+=0.365(5)$. This allows us to extract the amplitude $\xi_{0, {\rm 2nd}}^+$ of the second-moment correlation length appearing in eq.~(\ref{xi_amplitude}) by using the universal amplitude ratio $g_4^+$ \cite{PV-02}:
\begin{equation}
\label{g4}
g_4^+\equiv \frac{C_4^+}{(C^+)^2(\xi_{0, {\rm 2nd}}^+)^3},
\end{equation}
which corresponds to the critical value of the zero-momentum four-point coupling constant. From Ref.~\cite{CPRV-02} we quote $g_4^+=23.56(2)$. Using this value and the ones for the amplitudes $C^+$ and $C_4^+$ we finally obtain
\begin{equation}
\label{xi0_result}
\xi_{0, {\rm 2nd}}^+=0.415(2).
\end{equation}
Using the universal ratio $\xi_0^+/\xi_{0, {\rm 2nd}}^+=1.000200(3)$ \cite{CPRV-02}, from this one can calculate the non-universal amplitude of the true correlation length which, given the available precision, is identical to eq.~(\ref{xi0_result}). We have also analyzed directly the series for $\xi^2/\beta=m_2(\beta)/(\beta\chi(\beta))$. The corresponding result $\xi_{0, {\rm 2nd}}^+=0.419(7)$ appears to be less precise, but is in agreement with eq.~(\ref{xi0_result}).

The non-universal amplitude $B$ in eq.~(\ref{M_amplitude}) can obtained by using the universal amplitude-ratio $R_4^+$ \cite{PV-02}:
\begin{equation}
\label{R4}
R_4^+\equiv \frac{C_4^+B^2}{(C^+)^3}.
\end{equation}
From Ref.~\cite{CPRV-02} we quote $R_4^+=7.81(2)$. Using this value leads to
\begin{equation}
\label{B_result}
B=1.47(2).
\end{equation}
Finally, from eqs.~(\ref{xi0_result}) and (\ref{B_result}) together with $\beta/\nu=(1+\eta)/2=0.51819(8)$ \cite{CPRV-02} we obtain the coefficient appearing in eq.~(\ref{m_ansatz}):
\begin{equation}
B(\xi_0^+)^{\beta/\nu} = 0.933(6).
\end{equation}

\section{Monte Carlo simulations}
\label{sec:mc}
\Table{\label{mc_critical}The total MC steps $N_{\rm steps}$ and the steps $N_{\rm therm}$ disregarded for thermalization as used for the determination of the critical Casimir amplitude, for lattices with $L \ge 24$ and $L_\parallel=L/\rho$. Each step corresponds to $1$ Metropolis sweep and $L$ Wolff single-cluster flips. Runs marked with $^*$ have been split into independent runs with different random numbers and have then been recollected together: the number of steps reported there refers to the cumulated numbers.}
\begin{tabular}{@{}llll@{\hspace{3em}}lll}
\br
$L$  & $\rho$ & $N_{\rm steps}/10^3$ & $N_{\rm therm}/10^3$ & $\rho$ & $N_{\rm steps}/10^3$ & $N_{\rm therm}/10^3$ \\
\mr
$24$ & $1/6$  & $4000^*$             & $200$  & $1/8$  & $5100^*$             & $270$ \\
$32$ & $1/6$  & $5100^*$             & $300$  & $1/8$  & $5100^*$             & $270$ \\
$40$ & $1/6$  & \0$250$              & \0$20$ & $1/8$  & \0$250$              & \0$25$ \\
$48$ & $1/6$  & $1250^*$             & \0$80$ & $1/8$  & $1400^*$             & $140$ \\
$64$ & $1/6$  & $2400^*$             & $160$  & $1/8$  & $1400^*$             & \0$80$ \\
\mr
$24$ & $1/10$ & $2000$               & $100$   & $1/12$ & $1750$               & \0$20$ \\
$32$ & $1/10$ & $3280^*$             & $170$   & $1/12$ & $1020^*$             & \0$60$ \\
$40$ & $1/10$ & \0$100$              & \0\0$5$ & $1/12$ & \0\0$90$             & \0$20$ \\
$48$ & $1/10$ & \0$600^*$            & \0$30$  & $1/12$ & \0$450$              &\0$30$ \\
$64$ & $1/10$ & \0$540^*$            & \0$44$ &&& \\
\br
\end{tabular}
\endTable
\Table{\label{mc_profiles}The number of MC steps $N_{\rm steps}$, excluded thermalization, used for the determination of the order parameter profiles. Each step corresponds to $1$ Metropolis sweep and $L$ Wolff single-cluster flips. We have sampled the order parameter at $x/L=k/8$, with $k=-16,\ldots,16$, around the chemical step and for $z/L=j/8$, $j=1,\ldots, 7$. The coordinate system is chosen such that the confining surface with the spins fixed to form a chemical step corresponds to $z=0$, while the homogeneous surface with the spins fixed to $+1$ is located at $z=L-1$. In the lateral direction the origin $x=0$ corresponds, in the lower surface, to spins fixed to $+1$, the left neighbours of which are spins fixed to $-1$.}
\begin{tabular}{@{}lllll}
\br
$L$  & $\rho=1/6$           & $\rho=1/8$           & $\rho=1/10$           & $\rho=1/12$\\
     & $N_{\rm steps}/10^3$ & $N_{\rm steps}/10^3$ & $N_{\rm steps}/10^3$ & $N_{\rm steps}/10^3$ \\
\mr
$16$ & $300$                & $225$                & $180$                 & $150$  \\
$24$ & $200$                & $150$                & $120$                 & $100$  \\
$32$ & $100$                & $115$                & \0$90$                & \0$75$ \\
$40$ & $120$                & \0$90$               & \0$75$                & \0$60$ \\
$48$ & $100$                & \0$75$               & \0$60$                & \0$50$ \\
\br
\end{tabular}
\endTable

In this appendix we report certain technical details of the Monte Carlo simulations we have performed.
As explained in \sref{sec:simulations}, the evaluation of the Casimir force is carried out in two steps. First, we determine the thermal average $\<{\cal H}_2-{\cal H}_1\>_\lambda$ which appears in eq.~(\ref{I_def}). This is done by a standard Monte Carlo simulation for the ensemble given by the crossover Hamiltonian ${\cal H}_\lambda$ defined in eq.~(\ref{crossover_H}). We implement a combination of the standard Metropolis and Wolff cluster algorithms: each MC step consists of $1$ Metropolis sweep over the entire lattice in lexicographic order and $L$ Wolff single-cluster flips; $L$ denotes the slab thickness including the two surfaces of fixed spins, so that there are $L-2$ layers of fluctuating spins.
As random number generator we use the double precision SIMD-oriented Fast Mersenne Twister (dSFMT) \cite{Mersenne}. Some details of the simulations performed at the critical temperature are reported in \tref{mc_critical}. Our Simulations have been performed by using various clusters. At the critical point they took approximately 11.5 single-CPU years on a Intel Xeon\tm E5450 running at 3 Ghz, 11.5 single-CPU years on an IBM Power6\tm 575 running at 4.7 Ghz, 3 single-CPU years on a AMD Opteron\tm 852 running at 2.6 Ghz and 3 single-CPU years on a AMD Opteron\tm 248 running at 2.2 Ghz. The simulations off the critical point took approximately 38 single-CPU years on a Intel Xeon\tm E5450 running at 3 Ghz, 47 single-CPU years on an IBM Power6\tm 575 running at 4.7 Ghz and 6 single-CPU years on a AMD Opteron\tm 852 running at 2.6 Ghz.

The Metropolis update acts only on the fluctuating spins. Upon implementing the Wolff cluster algorithm, particular care has to be taken of the boundary spins, i.e., the fluctuating spins located at $z=1$ and $z=L-2$ which are connected to the two confining surfaces, where the spins are either fixed to $+1$ or to $-1$. The interactions of the latter ones with the surface is given by the surface contribution to the Hamiltonian
\begin{equation}
\label{Hsurface}
{\cal H}_{\rm S}=-\beta\sum_{i\in S_+}S_i + \beta\sum_{i\in S_-}S_i,
\end{equation}
where $S_+$ and $S_-$ are the lattice sites of the boundary spins which are connected to a surface spin fixed to $+1$ and $-1$, respectively. The presence of the interaction described by eq.~(\ref{Hsurface}) does not allow one to straightforwardly implement the Wolff algorithm. In this respect we introduce two fictious spins $S_p=1$ and $S_m=-1$ which correspond to the spins on the confining surfaces fixed to $+1$ and to $-1$, respectively, and we rewrite eq.~(\ref{Hsurface}) as
\begin{equation}
\label{Hsurface_full}
{\cal H}_{\rm S}=-\beta S_p\sum_{i\in S_+}S_i -\beta S_m\sum_{i\in S_-}S_i.
\end{equation}
We now want to promote $S_p$ and $S_m$ to actual fluctuating spins. In order to do so, we enlarge our phase space by adding those configurations of spins which correspond to reversed signs of $S_p$ and $S_m$, while mantaining the relative sign of $S_p$ and $S_m$ constant. In other words, we allow the boundary conditions on the confining surface to be flipped, so that the phase space is the union of the phase space corresponding to the geometry of \fref{lattice} and the phase space of the ``reversed'' geometry where the spins at the upper surface are fixed to $-1$ and those at the lower surface are fixed to $+1$ ($x<0$) or to $-1$ ($x\ge 0$). The complete partition function of this system reads:
\begin{eqnarray}
\label{completeZ}
\fl {\cal Z}=\sum_{\{\cal C'\}} \exp\left[\beta\left(\sum_{\<i j\>}S_i S_j + \sum_{i\in S_+}S_pS_i + \sum_{i\in S_-}S_mS_i \right)-\mu S_pS_m- D\sum_i S_i^2 \right],\nonumber\\
\mu\rightarrow +\infty,
\end{eqnarray}
where the sum is over the enlarged phase space $\{{\cal C}'\}=\{S_i=\pm 1,0\}\times\{S_m,S_p=\pm 1\}$ and an infinitely strong antiferromagnetic bond between $S_p$ and $S_m$ ensures that $S_p=-S_m$. The partition function given in eq.~(\ref{completeZ}) is now suitable for applying a cluster routine. At the beginning of the simulation we set $S_p=1$ and $S_m=-1$. The cluster routine selects randomly a spin among the fluctuating ones. If the spin is different from $0$ the cluster is expanded around such a spin. According to the partition function in eq.~(\ref{completeZ}), if the cluster reaches one of the confining surfaces, the surface spins are flipped and the expansion of the cluster continues for all the spins interacting with the confining surfaces.
This cluster move only acts on non-zero spins. Ergodicity is obtained by supplementing the dynamics with Metropolis sweeps. We mention that a full cluster algorithm is possible for the special value $D=\ln 2$ \cite{BLH-95}.
The critical Casimir force resulting from the partition function in eq.~(\ref{completeZ}) is the average of the force for the geometry of \fref{lattice} and the ``reversed'' geometry. Since the force in the two cases is identical, the result is indeed the desired critical Casimir force. In fact, upon inspecting eq.~(\ref{BC}) one realizes that the observable ${\cal H}_2-{\cal H}_1$, which appears in eq.~(\ref{I_def}), is invariant under spin flip.

The second step of the method consists of the numerical integration in eq.~(\ref{I_def}). For this purpose we employ the Gauss-–Kronrod quadrature formula with $15/7$ points \cite{KMN-89}. With this quadrature it is sufficient to sample the integrand in eq.~(\ref{I_def}) at $15$ points in $\lambda$, the positions of which are fixed. The integral can be estimated using the full set of sampled points or using a subset of $7$ points. By comparing the results of the numerical integration based on $15$ and $7$ points, respectively, we checked that the systematic error due to the discretization of the integral in eq.~(\ref{I_def}) is smaller than the statistical uncertainty of the Monte Carlo data.

The order parameter profiles presented in \sref{sec:profiles} have been obtained by a standard Monte Carlo simulation for the Hamiltonian given in eq.~(\ref{BC}). In \tref{mc_profiles} we report some details of these simulations. In this case a cluster flip which involve the confining surfaces changes the sign of the magnetization. Accordingly the correct profiles are obtained by keeping track of such flips.

\end{document}